\documentclass[aps,reprint,twocolumn,superscriptaddress,citeautoscript,amsmath,amssymb,floatfix,longbibliography,prl]{revtex4-2}
\usepackage{amsmath}
\usepackage{amssymb}
\usepackage{graphicx}
\usepackage{color}
\usepackage[colorlinks=true,linkcolor=blue,citecolor=blue]{hyperref}

\usepackage[resetlabels]{multibib}
\usepackage{siunitx}
\usepackage{amsfonts}
\usepackage{bm}
\usepackage{euscript}
\usepackage{textcomp}
\usepackage{gensymb}
\usepackage{float}
\usepackage{enumitem}
\usepackage{xcolor}
\usepackage{soul}
\usepackage[normalem]{ulem}

\begin{document}

\title{High-field magnetic phase diagrams of the {\it R}Mn$_6$Sn$_6$ kagome metals}

\author{Nil L}
\affiliation{Department of Physics and Astronomy, Iowa State University, Ames, Iowa, 50011, USA}

\author{Thais Victa Trevisan}
\affiliation{S\~{a}o Carlos Institute of Physics, University of S\~{a}o Paulo, PO Box 369, 13560-970, S\~{a}o Carlos, SP, Brazil.}

\author{R.~J.~McQueeney}
\affiliation{Department of Physics and Astronomy, Iowa State University, Ames, Iowa, 50011, USA}
\affiliation{Ames National Laboratory, Ames, Iowa, 50011, USA}

\date{\today}
 
\begin{abstract}
{\it R}Mn$_6$Sn$_6$ ($R=$~Gd-Lu) kagome metals are promising materials hosting flat electronic bands and Dirac points that interact with magnetism.  The coupling between the two magnetic $R$ and Mn sublattices can drive complex magnetic states with potential consequences for spin and charge transport and other topological properties.   Here, we use a detailed magnetic Hamiltonian to calculate and predict the magnetic phase diagrams for {\it R}Mn$_6$Sn$_6$ kagome metals within the mean-field approximation.  These calculations reveal a variety of collinear, noncollinear, and noncoplanar phases that arise from competition between various interlayer magnetic exchange interactions and magnetic anisotropies of the $R$ and Mn ions.  We enumerate these phases and their magnetic space groups for future analysis of their impact on topological and trivial bands near the Fermi surface.
\end{abstract}

\maketitle

\section{Introduction}

Magnetic topological materials have attracted great interest due to their ability to control bulk and surface electronic states through the breaking of time-reversal symmetry and other crystallographic symmetries.  For example, broken time-reversal symmetry is capable of generating novel surface states in topological insulators \cite{Chang2013,Chang2015,Liu2018} or bulk chiral electrons in Weyl semimetals \cite{Kiyohara2016,Liu2018}.  The emergence of spin chirality in complex helical magnetic states can also impart its own real-space topology on the transport and optical properties in topological materials.  In this respect, the family of {\it R}Mn$_6$Sn$_6$ ($R$166) compounds ($R=$~rare earth) with complex magnetism forming at high magnetic transition temperatures ($T_C\approx 400$ K) has been the subject of recent investigations \cite{Zhang2020, Yin:2020aa, Ma2021, Riberolles2022, Mielke2022, Kabir2022, Riberolles2023, Zeng2023, Lee2023, Riberolles2024a, Riberolles2024b}.

\begin{figure}[ht]
\includegraphics[width=0.8\linewidth,clip]{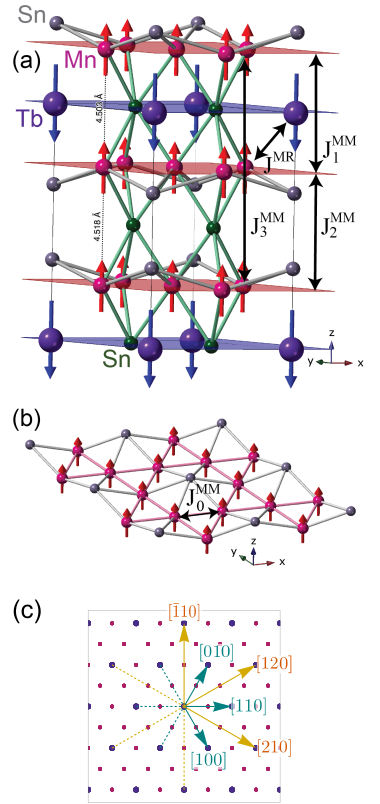} 
\caption{Crystal and magnetic structure of TbMn$_6$Sn$_6$. (a) Unit cell showing Tb and Mn magnetic sublattices forming the zero-field ferrimagnetic structure. (b) Depiction of a single Mn kagome layer with neighboring Sn layer. Magnetic interactions are labeled and described in more detail in the text. (c) Top view of the crystal depicting a single Mn kagome layer and a single Tb layer. In-plane high-symmetry directions $[uvw]=u\mathbf{a}+v\mathbf{b}+w\mathbf{c}$ are shown.}
\label{fig:structure}
\end{figure}

{\it R}166 compounds consist of bilayers of Mn kagome layers stacked along the crystallographic $\mathbf{c}$-axis and separated by triangular layers of  $R$ and Sn atoms, as illustrated in Fig.~\ref{fig:structure}. The kagome layers can naturally form flat electronic bands and Dirac points due to their special lattice geometry and may be manipulated by magnetic order.  As each Mn layer is ferromagnetic (FM), the spin-polarized Mn Dirac bands with large exchange splitting are candidates for generating a Chern insulator where Dirac points are gapped by spin-orbit coupling \cite{Xu2015,Yin:2020aa}. The Chern gap in a 2D kagome layer is largest for uniaxial Mn order.  However, Mn has an easy-plane magnetic anisotropy and competing exchange interactions between the Mn layers drives spiral antiferromagnetic (AF) order, as observed in YMn$_6$Sn$_6$ \cite{Venturini1996,Ghimire2020,Dally2021}.  

The introduction of magnetic rare-earths $R=$~Gd-Tm leads to a variety of magnetic phases caused by the strong magnetic anisotropy of the $R$ ion and also the antiferromagnetic coupling between $R$ and Mn layers \cite{ElIdrissi1991, Venturini1991, Venturini1992, Venturini1996}. TbMn$_6$Sn$_6$ has generated a great deal of interest since the uniaxial anisotropy of Tb provides ideal conditions for a 2D Chern insulator described above.  The tunability of the magnetic order for different rare-earths and as a function of field and temperature provides novel pathways for discovery of other topological phases \cite{Ma2021}.

Here, we investigate the complex magnetism in the family of $R$Mn$_6$Sn$_6$ kagome metals using a detailed magnetic model.  The model consists of magnetic exchange and single-ion anisotropy parameters informed by neutron diffraction \cite{Rosenfeld2008,Ghimire2020,Riberolles2024b}, inelastic neutron scattering (INS) \cite{Tils1998,Zhang2020,Riberolles2022,Riberolles2024a,Riberolles2024b}, and magnetization data \cite{Clatterbuck1999, Zaikov2000, Zajkov2000, Terentev2005,Kimura2006,Suga2006, Uhliova2006, Guo2007a,Guo2007b,Guo2008, Gorbunov2012}. Even though there may be some variation in these parameters across the rare-earth series, we use a single set of parameters for the entire family and employ Stevens scaling for the rare-earth magnetic anisotropy and empirical scaling for the Mn-$R$ exchange. This approach provides a window into the myriad of novel magnetic phases that are possible under identical crystal-field and exchange environments. 

Within a mean-field approach, we find the equilibrium magnetic states as a function of field and temperature, revealing a multitude of magnetic phases in good agreement with experimental data, where available.  These results justify the transferability of the magnetic interactions within the series and allow for a general understanding of the factors that control the magnetic phases. For example, these phases may be manipulated by thermal fluctuations of the $R$ magnetic moment that quench the $R$ anisotropy and reduce the effective Mn-$R$ exchange coupling.  As a consequence of this trend, all compounds evolve into easy-plane magnets at high temperatures and low field strength. For the heavier rare-earths with $R=$ Er and Tm,  relatively weak Mn-$R$ exchange leads to the appearance of complex spiral phases \cite{Malaman1999, Lefevre2003, Riberolles2024b} similar to YMn$_6$Sn$_6$.  Overall, we find many interesting phases and novel features in the phase diagrams (such as a liquid-gas-like critical point in DyMn$_6$Sn$_6$ and HoMn$_6$Sn$_6$).  We determine the magnetic space group for each phase, which can assist in future investigations of the impact of these magnetic phases on kagome electronic bands.

\section{Magnetic Hamiltonian}

The $R$166 compounds consist of $R$ local moments and Mn moments that have an metallic character. To analyze the magnetism, we treat the Mn moments also as local moments with an effective spin $s=1$ and effective $g$-factor of $g=2.17$ that reproduces the correct magnetic moment and its coupling to external magnetic fields.  Treatment of Mn as local moments is an approximation that is well-justified for our purposes. For example, local moment models similar to the one we describe below have been used to predict critical temperatures and fields for various for $R$166 compounds \cite{Zaikov2000,Zajkov2000,Terentev2005,Guo2007a,Guo2007b,Guo2008,Rosenfeld2008,Ghimire2020,Riberolles2024b}, and accurately describes the magnon excitations up to energies of approximately 100 meV \cite{Zhang2020,Riberolles2022, Riberolles2024a, Huang2024}. Furthermore, there is no evidence that the magnitude of the Mn moment is dependent on temperature or moment direction, as is common for weak itinerant magnets.

We analyze a magnetic Hamiltonian that is comprised of isotropic exchange interactions ($\mathcal{H}_{ex}$) between Mn-$R$ and Mn-Mn ions which are presumably mediated by some combination of local moment (superexchange) and itinerant (RKKY) couplings. The Hamiltonian also includes anisotropic single-ion crystalline electric field (CEF) terms summed over all $R$ ($\mathcal{H}_A^{R}$) and Mn sites ($\mathcal{H}_A^{\rm Mn}$), and Zeeman terms for applied fields ($\mathcal{H}_{Z}^R$) and ($\mathcal{H}_{Z}^{\rm Mn}$). The full Hamiltonian is
\begin{equation}
\mathcal{H}=\mathcal{H}_{ex}+\mathcal{H}_{A}^{\rm Mn}+\mathcal{H}_{Z}^{\rm Mn}+\mathcal{H}_A^R+ \mathcal{H}_{Z}^R
\label{eqn:Hamiltonian}
\end{equation}

The exchange Hamiltonian is given by
\begin{equation}
\mathcal{H}_{\rm ex}=\frac{1}{2}\sum_{i,j} \mathcal{J}^{MM}_{ij} \mathbf{s}_i \cdot \mathbf{s}_j + \frac{1}{2}\mathcal{J}^{MR} \sum_{i,j} \mathbf{s}_i \cdot \mathbf{S}_j
\label{eqn:Hex}
\end{equation}
where $\mathcal{J}^{MM}_{ij}$ represent various, and quite strong, intralayer and interlayer magnetic couplings between Mn spins (${\bf s}$). Here, $\mathcal{J}^{MR}>0$ is the AF coupling between neighboring Mn and $R$ spins (${\bf S}$) as shown in Fig.~\ref{fig:structure}. The $\mathcal{J}^{MM}$ terms are obtained from neutron data for TbMn$_6$Sn$_6$ \cite{Riberolles2022, Riberolles2023} and ErMn$_6$Sn$_6$ \cite{Riberolles2024b} and we approximate that Mn-Mn couplings are independent of $R$.  Some ambiguity exists in the literature regarding in the relative signs of the competing $\mathcal{J}^{MM}_1$ and $\mathcal{J}^{MM}_3$ interactions that result in spiral phases \cite{Ghimire2020,Riberolles2023,Riberolles2024b}. Here we use $\mathcal{J}^{MM}_3>0$ (AF) and $\mathcal{J}^{MM}_1<0$ (FM) which is consistent with fits to INS data.  This choice has no effect on the phase diagrams presented here, but may affect the behavior of spiral phases in planar magnetic fields. From both experimental \cite{Tils1998, Riberolles2022, Riberolles2023, Riberolles2024b} and theoretical works \cite{Lee2023}, $\mathcal{J}^{MR}$ is known to decrease in approximately linearly from Gd (2.0 meV) to Tm (1.19 meV) and we adopt this trend.  First-principles calculations associate this decrease with a reduction of the 4$f$ and 5$d$ orbital overlap  \cite{Lee2023}.

For Mn, a simple easy-plane anisotropy is employed, consistent with the planar ferrimagnetic and helical ground states of GdMn$_6$Sn$_6$ and YMn$_6$Sn$_6$, respectively.  A single Mn site ($i$) has an anisotropy term given by
\begin{equation}
(\mathcal{H}_{A}^{\rm Mn})_i=D^Ms_{iz}^2
\label{eqn:HAM}
\end{equation}
Various values of $D^M$ are reported in the literature \cite{Rosenfeld2008, Gorbunov2012, Ghimire2020, Riberolles2023} and here we use an average value of $D^M=0.26$ meV.


For magnetic rare-earth ions, a more complex anisotropic behavior arises due to the orbital 4$f$ states of $R$ in the crystalline electric field (CEF) potential of neighboring ions. The local $R$ Hamiltonian for a single site is given by
\begin{equation}
(\mathcal{H}_A^R)_i=\sum_{lm}B_l^m\mathcal{O}_l^m({\bf J}_i)
\label{eqn:HAR}
\end{equation}
where $B_l^m$ are the CEF parameters, $\mathcal{O}_l^m$ are the Stevens operators defined in terms of the total angular momentum  ${\bf J}={\bf S}/(g_R-1)$ where $g_R$ is the Land\'{e} $g$-factor of the rare-earth.  For the $D_{6h}$ point symmetry of the $R$ ion, only  $B_2^0$, $B_4^0$, $B_6^0$, and $B_6^6$ are nonzero.  

The $B_l^m$ for different $R$ ions is determined by the equation 
\begin{equation}
B_l^m=\langle r^l \rangle \theta_l A_l^m
\label{eqn:Stevens}
\end{equation}
where $A_l^m$ is the strength of the different multipolar components of the crystal field, $\theta_l$ is the Stevens factor and the $\langle r^l \rangle$ is the average $l^{th}$ radial moment of the 4$f$ electron cloud of the $R$ ion.  We assume that the CEF potential is constant across the series with approximate values of $A_2^0 = 5.33~{\rm meV}~a_0^{-2}$, $A_4^0 = -6.6~{\rm meV}~a_0^{-4}$, $A_6^0 \approx 0$, and  $A_6^6 = 1.38~{\rm meV}~a_0^{-6}$ (where $a_0$ is the Bohr radius). Other parameters of Eq.~\ref{eqn:Stevens} and the generated CEF parameters are shown in Table \ref{tbl:CEF}.

The Zeeman energies for a single site on each sublattice are determined by the applied magnetic field ($\mu_0{\bf H}$).
\begin{equation}
    (\mathcal{H}_Z^R)_i=-g_R\mu_B{\bf J}_i\cdot {\mu_0\bf H}
\label{eqn:HZR}
\end{equation}
and
\begin{equation}
    (\mathcal{H}_Z^{\rm Mn})_i = -g\mu_B{\bf s}_i\cdot \mu_0{\bf H}
\label{eqn:HZM}
\end{equation}
where $g\approx 2.17$ is the Mn $g-$factor chosen to reproduce the measured magnetic moment.

\begin{table*}
\caption {Magnetic Hamiltonian parameters for $R$166 compounds used in mean-field calculations.  We also include $D^M=0.26$ meV, $\mathcal{J}^{MM}_0=-28.8$ meV, $\mathcal{J}^{MM}_1=-4.4$ meV, $\mathcal{J}^{MM}_2=-19.2$ meV, $\mathcal{J}^{MM}_3=2.3$ meV parameters of the Mn sublattice.}
\renewcommand\arraystretch{1.25}
\centering
\begin{tabular}{ c | c | c | c | c | c | c  }
\hline\hline
$R$ ion							&Gd   & Tb 		& Dy 		& Ho 		& Er 			& Tm 		\\
$ (L,S,J)$						&(0,$\frac{7}{2}$,$\frac{7}{2}$)& $(3,3,6)$	& (5,$\frac{5}{2}$,$\frac{15}{2}$) 	& (6,2,8) 		& (6,$\frac{3}{2}$,$\frac{15}{2}$) 	& (5,1,6)	\\
\hline					
$\langle r^2 \rangle$ ($a_0^2$)		&&~0.829~		&~0.790~		&~0.756~		& ~0.724~		&~0.695~ 			\\ 	
$\langle r^4 \rangle$ ($a_0^4$) 	&& ~1.68~		&~1.55~ 		&~1.43~		&~1.33~		& ~1.24~			\\
$\langle r^6 \rangle$ ($a_0^6$)		&& ~6.91~		& ~6.21~		& ~5.63~		& ~5.15~		& ~4.72~ 			\\
$\theta_2$ ($10^{-2}$)			&&-1.01		&-0.635		&-0.222		& 0.254		& ~1.01~ 			\\
$\theta_4$  ($10^{-4}$) 		&	&1.22		&-0.592		&-0.333		&0.444		&2.07			\\
$\theta_6$  ($10^{-6}$)			&&-1.12 		&1.035 		&-1.29 		&2.07 		&5.61 			\\
$B_2^0$ (meV)					&0& -4.46E-2 	& -2.67E-2		& -8.95E-3		& 9.80E-3		& 3.74E-2				\\
$B_4^0$ (meV)					&0& -1.36E-3  	& 6.06E-4 		& 3.14E-4		& -3.90E-4		& -1.34E-3			\\
$B_6^0$ (meV)					&0& 0 & 	0		& 0		& 0		& 0		\\
$B_6^6$ (meV)					&0& -1.07E-5	 	& 8.87E-6		& -1.01E-5		& 1.47E-5 		&3.65E-5 				\\
$K_1$ (meV)					&0& 44.73		& -45.39		& -32.71		& 30.38		& 35.96		 	\\
$K_2$ (meV)					&0& -35.27		&  43.40		&  30.03		& -27.93		& -34.71			\\
$K_3$ (meV)					&0& 0 & 	0		& 0		& 0		& 0		\\
$K_3'$ (meV)				&0	& -0.11		& 0.50		& -0.91		& 0.83		& 0.38				\\
$\mathcal{J}^{MR}$~(meV)          & 2.0   & 1.83  &1.67   &1.51   &1.35   &1.19    \\
\hline\hline
\end{tabular}
\label{tbl:CEF}
\end{table*}

In Table \ref{tbl:CEF}, we also report the classical magnetic anisotropy energy (MAE) constants $K_1$, $K_2$, $K_3$ and $K_3'$ for the $R$ ion which are useful for interpreting the resulting magnetic phase diagrams. The MAE constants are related to the CEF parameters according to the following relations 
\begin{align}
&K_1=-3J^{(2)}B_2^0-40J^{(4)}B_4^0-168J^{(6)}B_6^0 \nonumber \\
&K_2=35J^{(4)}B_4^0+378J^{(6)}B_6^0 \nonumber \\
&K_3=-231J^{(6)}B_6^0 \nonumber \\
&K_3'=J^{(6)}B_6^6
\label{eqn:MAE_const} 
\end{align}
where $J^{(2)}=J(J-\frac{1}{2})$, $J^{(4)}=J^{(2)}(J-1)(J-\frac{3}{2})$, and  $J^{(6)}=J^{(4)}(J-2)(J-\frac{5}{2})$. We can then write the classical MAE for the $R$ ion as
\begin{align}
\nonumber
    E_A^R &= K_1\sin^2\theta^R + K_2\sin^4\theta^R+K_3\sin^6\theta^R \nonumber \\ 
    & + K_3'\sin^6\theta^R\cos6\varphi^R
\end{align}
where $\theta^R$ and $\varphi^R$ are the spherical angles defining the direction of the $R$ moment.

\section{Mean-field description of the free energy}

This section describes the details of our constrained self-consistent mean-field approach.  Those interested only in the results may skip this section.  We start with a mean-field decomposition of the exchange Hamiltonian Eq.~\ref{eqn:Hex} given by
\begin{align}
&(\mathcal{H}_{ex})_{MF} = \frac{1}{2}\mathcal{J}^{MR} \sum_{i,j} (\langle{\bf s}_i \rangle \cdot {\bf S}_j  + {\bf s}_i \cdot \langle {\bf S}_j\rangle - \langle {\bf s}_i \rangle \cdot \langle {\bf S}_j\rangle) \nonumber \\
&+ \frac{1}{2}\sum_{i,j} \mathcal{J}^{MM}_{ij} (\langle {\bf s}_i\rangle \cdot {\bf s}_j + {\bf s}_i \cdot \langle {\bf s}_j \rangle -\langle {\bf s}_i \rangle \cdot \langle {\bf s}_j \rangle).
\label{eqn:Hex_MF}
\end{align}
We use these terms to generate local Hamiltonians for $R$ and Mn sites which include the molecular and applied fields
\begin{equation}
\label{eqn:HR_MF}
(\mathcal{H}_{MF}^R)_i = (\mathcal{H}_A^R)_i - g_R\mu_B{\bf J}_i\cdot[\mu_0{\bf H} + ({\bf B}^R)_i]
\end{equation}
and
\begin{equation}
\label{eqn:HM_MF}
(\mathcal{H}_{MF}^{\rm Mn})_i = (\mathcal{H}_A^{\rm Mn})_i - g\mu_B{\bf s}_i\cdot [\mu_0{\bf H} + ({\bf B}^{\rm Mn})_i]
\end{equation}
 Here, $({\bf B}^R)_i$ and $({\bf B}^{\rm Mn})_i$ represent the self-consistently determined molecular fields acting on each $R$ and Mn ion, respectively.

The self-consistent solution to the mean-field Hamiltonian at some applied field and temperature ($T$) is obtained by minimization of the free energy, $\mathcal{F}=\mathcal{F}_{ex}+\mathcal{F}^{R}+\mathcal{F}^{\rm Mn}$. Here, $\mathcal{F}^{R}$ and $\mathcal{F}^{\rm Mn}$ are the free energies of the local crystal field problem in the combined applied and molecular field given in Eqs.~\ref{eqn:HR_MF} and \ref{eqn:HM_MF}. A general solution to this problem is quite difficult, so we make certain approximations that are justified for the $R$166 compounds. 

First, all $R$166 compounds have strong FM intralayer Mn-Mn interactions ($\mathcal{J}_0^{MM}$). Thus, each Mn or $R$ layer is FM and we assume that moment directions can only vary from layer-to-layer and not within a layer. We define $\langle{\bf s}_{k\ell}\rangle$ as the average Mn spin in layer $k=1,2$ of unit cell $\ell$ and $\langle{\bf S}_{\ell}\rangle$ as the average $R$ spin in unit cell $\ell$. The molecular fields are
\begin{equation}
({\bf B}^{R})_{\ell}=-\frac{6\mathcal{J}^{MR}(g_R-1)}{g_R\mu_B}(\langle{\bf s}_{1\ell}\rangle + \langle{\bf s}_{2\ell}\rangle)
\label{eqn:BR_gen}
\end{equation}
and 

\begin{align}
({\bf B}^{\rm Mn})_{k\ell} &= -\frac{1}{g\mu_B}\Big[2\mathcal{J}^{MR} \langle{\bf S}_{\ell}\rangle + 4\mathcal{J}^{MM}_{0} \langle{\bf s}_{k\ell} \rangle \nonumber \\
&+\mathcal{J}^{MM}_{1} \langle {\bf s}_{k'\ell}\rangle +\mathcal{J}^{MM}_{2} \langle{\bf s}_{k'\ell+(-1)^k}\rangle \nonumber \\ 
&+2\mathcal{J}^{MM}_{3} (\langle{\bf s}_{k\ell-1}\rangle + \langle{\bf s}_{k\ell+1}\rangle)\Big]
\label{eqn:BM_gen2}
\end{align}
where $k\neq k'$ and $\mathcal{J}^{MM}_{n}$ label the intralayer and interlayer Mn-Mn interactions shown in Fig.~\ref{fig:structure} with values given in the caption of Table \ref{tbl:CEF}.

We have used two different sets of constraints to evaluate the equilibrium structures.  In the {\it coplanar model}, we assume that $R$ and Mn each form FM sublattices in which the moments can have a relative canting angle between them.  We note that we have also studied a less constrained model where the magnetization of the two Mn sublattices within the unit cell are unconstrained, but investigations of this more complicated model did not reveal any new phases.  In the {\it spiral model}, regular layer-to-layer rotations of the $R$ and Mn moments are allowed to form a regular spiral or conical spiral.  We are able to investigate this model only under applied fields pointing along the spiral propagation vector (i.e., along the $c$-axis).

{\it Coplanar model}. For uniaxial (Tb), easy cone (Dy, Ho), and easy-plane (Gd) anisotropies, only $q=0$ coplanar (canted) or collinear ferrimagnetic (FIM) phases have been reported. In this approximation, all Mn moments point in the same direction  $\langle{\bf s}_{k\ell} \rangle = \langle{\bf s}_{k'\ell'}\rangle = \langle{\bf s}\rangle$. The independent direction of the $R$ moment $\langle{\bf S}\rangle$ defines a plane containing both Mn and $R$ moment vectors.  The molecular fields for all $R$ sites and for all Mn sites simplify to
\begin{equation}
{\bf B}^{R}=-\frac{12\mathcal{J}^{MR}(g_R-1)}{g_R\mu_B}\langle{\bf s}\rangle
\label{eqn:BR_co}
\end{equation}
\begin{align}
{\bf B}^{\rm Mn} &= -\frac{1}{g\mu_B}\Big[2\mathcal{J}^{ME} \langle{\bf S}\rangle + (4\mathcal{J}^{MM}_{0} +\mathcal{J}^{MM}_{1} \nonumber \\
&+\mathcal{J}^{MM}_{2}+2\mathcal{J}^{MM}_{3})\langle{\bf s}\rangle\Big]
\label{eqn:BM_co}
\end{align}

Finding the equilibrium structure requires minimizing the free energy while varying the two vectors $\langle{\bf s}\rangle$ and $\langle{\bf S}\rangle$.  We find $\mathcal{F}^R$ and $\mathcal{F}^{\rm Mn}$ by solving the CEF problem at each $T$ and ${\bf H}$ for a set of four spherical angles $[\theta^R,\varphi^{R},\theta^{\rm M},\varphi^{\rm M}]$ and determining the free energy through the partition function; $\mathcal{F}^R=-k_BT {\rm ln}(Z^R)$ and $\mathcal{F}^{\rm Mn}=-6k_BT {\rm ln}(Z^{\rm Mn})$.  The spin magnitudes are determined self-consistently, defining the spin vectors

\begin{align}
   & \langle s_x \rangle = \langle s \rangle\sin\theta^M \cos\varphi^M \nonumber \\
    & \langle s_y \rangle = \langle s \rangle\sin\theta^M \sin\varphi^M \nonumber \\
    &\langle s_z \rangle = \langle s \rangle\cos\theta^M 
\label{eqn:Mn_mom}
\end{align}
and
\begin{align}
   & \langle S_x \rangle = \langle S \rangle\sin\theta^R \cos\varphi^R \nonumber \\
    & \langle S_y \rangle = \langle S \rangle\sin\theta^R \sin\varphi^R \nonumber \\
    &\langle S_z \rangle = \langle S \rangle\cos\theta^R
\label{eqn:R_mom}
\end{align}

The exchange contributions to the free energy are evaluated as
\begin{align}
\mathcal{F}_{ex} &= - 3\langle s\rangle^2(4\mathcal{J}^{MM}_{0} +\mathcal{J}^{MM}_{1}+\mathcal{J}^{MM}_{2}+2\mathcal{J}^{MM}_{3}) \nonumber \\
&-12\mathcal{J}^{MR} \langle{\bf s}\rangle \cdot \langle{\bf S}\rangle.
\label{eqn:Fex_co}
\end{align}
A nonlinear minimization procedure is used to find the equilibrium values of the spin angles and their magnitudes.  The magnetic phases encountered in these studies are shown in Table~\ref{tbl:phases} and described in more detail below.

\begin{table*}
\centering
\caption{Different collinear, canted, noncollinear, and noncoplanar phases are encountered in mean-field calculations for $R$Mn$_6$Sn$_6$ compounds. Phases are indicated by the following abbreviations: ferrimagnetic (FIM), forced ferromagnetic (FF), vertical plane canted (VP), horizontal plane canted (HP), and vertical conical spiral (VCS).  The letter following the phase name indicates that the moment direction or canting plane contains a symmetry axis or plane labeled as $a$, $b$, and $c$ which correspond to the [100], [120], and [001] directions in hexagonal notation, respectively. The $\pm$ label for canted phases corresponds to whether the projection of the $R$ moment onto the field direction is parallel ($+$) or antiparallel ($-$) to the field.  Magnetic space groups are listed in green text.  FF and FIM phases with parallel or antiparallel Mn and $R$ sublattice magnetizations, respectively, have the same magnetic space group.  FIM-tilt phases are special cases of the VP-canted structure with the canting angle of $\pi$ and have the same magnetic space group.}
\includegraphics[width=\textwidth]{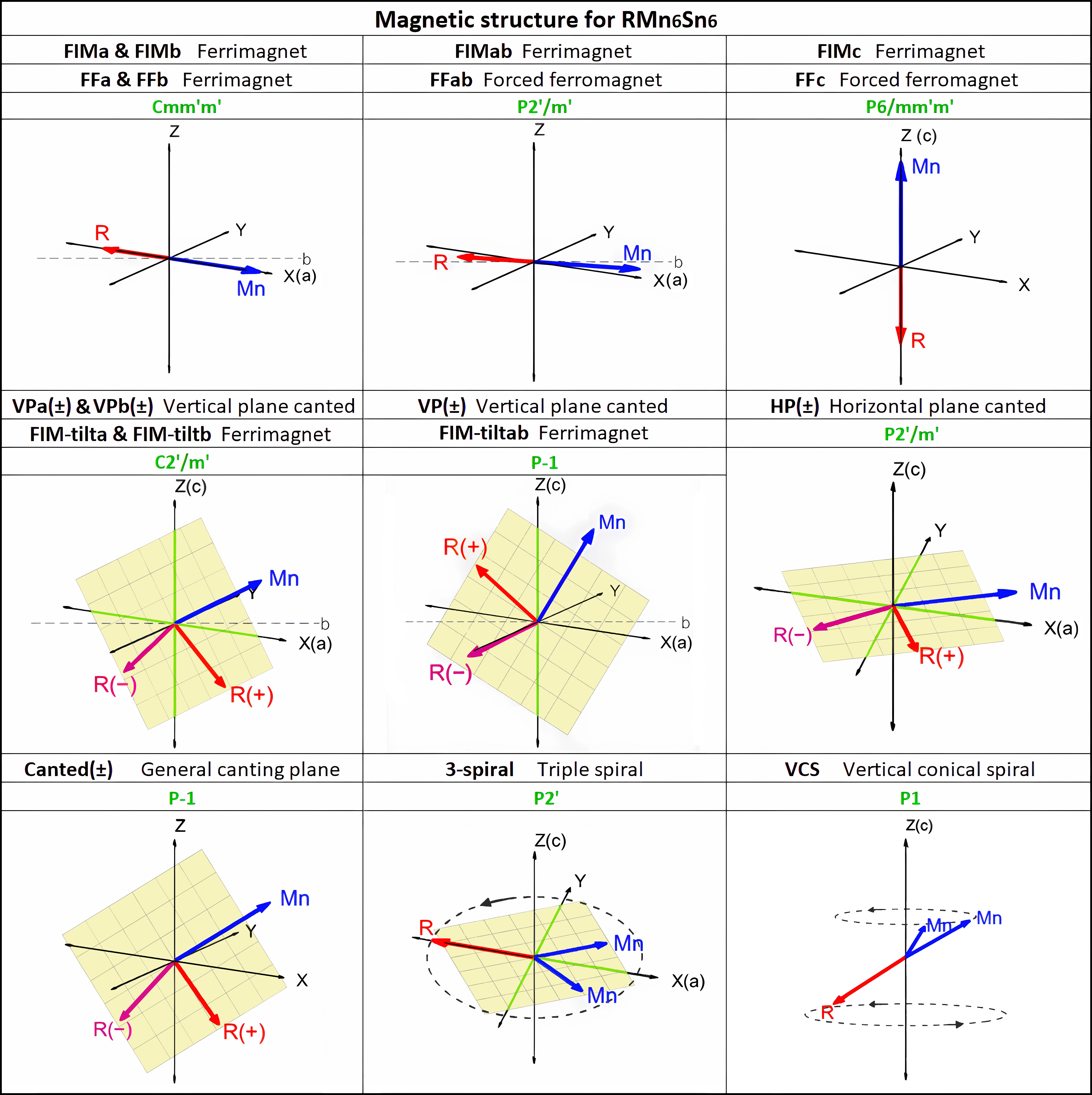}
\label{tbl:phases}
\end{table*}

{\it Spiral model}. 
$R$166 compounds with a weak Mn-$R$ interaction and planar anisotropy ($R=$ Y, Er, Tm) lead to noncollinear and noncoplanar (spiral, conical spiral, and fan) phases with chiral properties at low fields. For easy-plane ($B_6^6=0$) configurations in zero field, only the relative angles between spins in each layer factor in to the free energy. The periodicity is defined by the angle $\Phi$ between like layers in adjacent cells (coupled by $\mathcal{J}^{MM}_3$) and the angle $\delta$ is the rotation of the Mn spins between strongly coupled layers (through $\mathcal{J}^{MM}_2$).

In the general case of arbitrary applied field direction and nonzero planar anisotropy ($B_6^6$), layer-to-layer variations of the moments can form quite complex noncollinear distorted spiral, fan, and cycloid-like structures.  There is strong interest in these phases due to evidence for novel transport properties.  For example, observation of the topological Hall effect in Y166 and Er166 in applied planar fields suggests the development of a nonzero scalar spin chirality \cite{Ghimire2020,Fruhling2024}.  Solving the self-consistent mean-field problem requires the definition of unique molecular fields for each Mn and $R$ layer extended over multiple unit cells. We leave this for future study.

For a vertical magnetic field, the situation is much simpler since we nominally maintain the spiral angles $\delta$ and $\Phi$ and introduce the canting angles of the $R$ and Mn sublattices $\theta^R$ and $\theta^M$.  We define a coordinate system that rotates with the spiral. We set the $R$ moment to point towards the $x$-axis in the rotating coordinate system while bisecting the angle ($\Phi-\delta$) between the Mn moment directions in the layers above and below. The moment directions in the rotating system are

\begin{align}
\label{eqn:spins_spiral}
\hat{\bf S} &= -{\rm sin} \theta^R\hat{x} -{\rm cos} \theta^R \hat{z} \\
\hat{\bf s}_{1\ell} &= {\rm sin} \theta^M {\rm cos} \Big(\ell\Phi-\frac{\Phi-\delta}{2}\Big)\hat{x} \nonumber \\
&+{\rm sin} \theta^M {\rm sin} \Big(\ell\Phi-\frac{\Phi-\delta}{2}\Big)\hat{y} +{\rm cos} \theta^M \hat{z} \\
\hat{\bf s}_{2\ell} &= {\rm sin} \theta^M {\rm cos} \Big(\ell\Phi+\frac{\Phi-\delta}{2}\Big)\hat{x} \nonumber \\
&+{\rm sin} \theta^M {\rm sin} \Big(\ell\Phi+\frac{\Phi-\delta}{2}\Big)\hat{y} +{\rm cos} \theta^M \hat{z}
\end{align}
The corresponding molecular fields in the rotated system are given by Eq.~\ref{eqn:BR_gen} and \ref{eqn:BM_gen2} and the contribution of the exchange to the free energy is
\begin{align}
\label{eqn:Fex_spiral}
&\mathcal{F}_{ex} =  -3\langle s\rangle^2 \Big [4\mathcal{J}^{MM}_0 \nonumber \\
&+ \mathcal{J}_1^{MM} ({\rm sin}^2\theta^M {\rm cos}(\Phi-\delta)+{\rm cos}^2\theta^M) \nonumber \\
& + \mathcal{J}_2^{MM}  ({\rm sin}^2\theta^M {\rm cos}\delta+{\rm cos}^2\theta^M) \nonumber \\
&+ 2\mathcal{J}_3^{MM} ({\rm sin}^2\theta^M {\rm cos}\Phi+{\rm cos}^2\theta^M)\Big] \nonumber \\ 
&+12\mathcal{J}^{MR} \langle S\rangle \langle s\rangle [{\rm sin} \theta^M {\rm sin} \theta^R  {\rm cos} \Big(\frac{\Phi-\delta}{2}\Big) +{\rm cos} \theta^M {\rm cos} \theta^R]
\end{align}

We note that there is overlap between the coplanar and spiral models. For example, the spiral model can produce a subset of collinear and coplanar phases depending on the values of $\theta^M$ and $\theta^R$ when $\Phi,\delta = 0$ or $\pi$.  In cases where the equilibrium state is one of these phases, both coplanar and spiral models yield the same result. 

{\it Calculation of magnetization}.  
The outcome of the energy minimization is a set of angles in either the coplanar or spiral model, as well as the self-consistent magnitude of the sublattice spin (magnetic moment). In the coplanar model, the net magnetization per unit cell is given by reference to Eqs.~\ref{eqn:Mn_mom} and \ref{eqn:R_mom}
\begin{equation}
M_{\alpha} = 6g\mu_B\langle s_{\alpha}\rangle + g_R\mu_B \langle S_{\alpha}\rangle/(g_R-1)
\end{equation}
In the spiral phases encountered in the spiral model, the circulation of moments in the $xy$-plane means that the net magnetization can only point in the $z$-direction. This is given by
\begin{equation}
M_{z} = 6g\mu_B\langle s\rangle \cos\theta^M + g_R\mu_B \langle S\rangle \cos\theta^R/(g_R-1)
\end{equation}
Experimental data typically report the net magnetization component along the applied field direction ${\bf M} \cdot {\hat{\bf H}}=\sum_{\alpha}M_{\alpha}\hat{H}_{\alpha}$.

\vspace{4\baselineskip}

\section{Magnetic States and Symmetries}

The $R$166 family of kagome magnets belongs to the space group P6/mmm (No. 191), a symmorphic space group generated by a threefold rotation around $\hat{z}$ ($C_{3z}$, where $\hat{z}$ is parallel to the crystallographic $\mathbf{c}$ axis), a twofold rotation also around $\hat{z}$ ($C_{2z}$), a twofold rotation around the crystallographic direction $[100]\equiv \mathbf{a}_1$ ($C_{2,[100]}$), and spatial inversion ($\mathcal{P}$). Any $R$ atom can be chosen as the inversion center of the crystal. Here, we describe the magnetic structures that are encountered and their magnetic space group (MSG), as shown in Table \ref{tbl:phases}. In what follows, we make the definition that $a$ describes the [1,0,0] hexagonal direction and $b$ describes the inequivalent [1,2,0] hexagonal direction, as shown in Fig.~\ref{fig:structure}(c).

In the paramagnetic phase, the Mn and $R$ moments are disordered and the spatial distribution of the expected values of the magnetic moments remains unchanged when transformed by the symmetries of the crystal and their combination with time-reversal symmetry ($\mathcal{T}$). Therefore, the paramagnetic state is invariant under the symmetry operations of the magnetic gray group P6/mmm1$'$ (No. 191.234). 

When the magnetic moments order, time-reversal symmetry and some of the crystallographic symmetries are broken. However, certain combinations of time-reversal symmetry and spatial symmetries in P6/mmm1$’$ might leave the magnetic state invariant. We determine the magnetic space group of each identified magnetic state by finding the subset of operations of P6/mmm1$’$ that leave the distribution of magnetic moments unchanged. Since all ordered states considered in this work consist of magnetic moments that are ferromagnetically aligned in each $R$ and Mn layer, we can focus on the moment directions as a function of the $z$-coordinate of the layer. This greatly simplifies the analysis as the problem reduces to finding the symmetries of a one-dimensional chain of magnetic moments along $z$. Also, none of the symmetries of the parent crystallographic group (and, therefore of the parent gray group) maps an $R$ layer into an Mn layer and vice versa. Therefore, we can treat the $R$ moments and the Mn moments independently in our symmetry analysis. In other words, we first identify the symmetry groups that leave each of these moments invariant independently and the intersection of these two sets yields the MSG of the system. 

{\it Collinear FIM and FF states}. 
For collinear states, the set of operations that are preserved by the $R$ moments is identical to that of the Mn moments. We thus focus on the symmetry properties of only one of them. Besides, these symmetries are insensitive to whether the $R$ and Mn moments are parallel or antiparallel. Therefore, the two types of collinear states considered in this work, the ferrimagnetic (FIM) and forced ferromagnetic (FF) states, have the same symmetries. Consider first the Mn moment pointing along $\mathbf{c}$ (FIM$c$ or FF$c$). In these states, rotations around $\hat{z}$ are preserved. Spatial inversion ($\mathcal{P}$) is also preserved because the magnetic moments in the Mn layers related by $\mathcal{P}$ are identical. Moreover, twofold rotations around an axis in the $ab$-plane are generically broken in the FIM$c$ and FF$c$ states since they reverse the magnetic moments. However, time-reversal ($\mathcal{T}$) combined with these rotations recover the original moment orientation. As a consequence, the subset of the parent gray group that leaves FIM$c$ and FF$c$ states invariant is generated by $C_{2z}$, $C_{3z}$, $\mathcal{P}$ and $\mathcal{T}C_{2,[100]}$, and corresponds to the MSG P6/mm$'$m$'$ (No. 191.240).

We can also have FIM and FF states with Mn moment purely in the $ab$-plane. In this case, $C_{3z}$ and $C_{2z}$ are broken, but some of the in-plane twofold rotations present in the gray group might be preserved depending on the moment orientation. There are two distinct situations. First, if the Mn moments point along one of the in-plane high-symmetry directions illustrated in Fig.\ref{fig:structure}(c), then, besides inversion, a twofold rotation around this direction and a combination of a twofold rotation around the axis orthogonal to this direction and time-reversal are symmetries of the magnetic state. Time-reversal combined with twofold rotation around $\hat{z}$ is also preserved. For concreteness, in the case of a FIM$a$ state (see Table \ref{tbl:phases}), the MSG is generated by $\mathcal{P}$, $C_{2,[100]}$ and $\mathcal{T}C_{2,[120]}$, and corresponds to Cmm$'$m$'$ (No. 65.486). The MSG is the same for $ab$-plane FIMs pointing along the six in-plane high-symmetry [100]-directions and also the six [120]-directions (FIM$b$). The second case corresponds to Mn moment pointing in an arbitrary (nonsymmetry) direction of the plane (FIM$ab$). Then all rotations other than $\mathcal{T}C_{2z}$ are broken and the MSG lowers to P2$'$/m$'$ (No. 10.46), generated by identity $\mathcal{P}$ and $\mathcal{T}C_{2z}$. 

The combination a FIM$c$ and a FIM state in the $ab$-plane gives a tilted FIM state. Similarly, a combination between an FF$c$ and a FF state in the $ab$-plane gives a tilted FF structure. In these states, a finite $z$-component of the moments breaks the in-plane rotations which are not combined with time reversal. It also breaks $\mathcal{T}C_{2z}$. Therefore, tilting an in-plane FIM or FF where the component of the Mn moment in the $ab$-plane points in a high-symmetry direction (FIM-tilt$a$ or FIM-tilt$b$), lowers the MSG from Cmm$'$m$'$ to C2$'$/m$'$ (No. 12.62). Besides, tilting an in-plane FIM or FF with an $ab$-plane component of the Mn moment in an arbitrary direction (FIM-tilt$ab$) lowers the MSG P2$'$/m$'$ to P$\bar{1}$ (No. 2.4).  

{\it Canted states}. 
We move now to the symmetry analysis of the coplanar canted states. There are three categories of canted states. The first corresponds to the vertical plane canted (VP) structures. In these states, $R$ and Mn moments point along unrelated directions within a plane spanned by $\hat{z}$ and a direction in the $ab$-plane. Since they share either parallel (denoted by $+$ in the state name) or anti-parallel ($-$) $ab$-plane components, their MSG is the same as the tilted FF and tilted FIM. Thus, we designate VP$a(\pm)$, VP$b(\pm)$, and VP$ab(\pm)$ as vertical plane canted structures.

The second category corresponds to the horizontal plane canted structures, HP($\pm$). In these structures, although both Mn and $R$ moments are purely in the $ab$-plane, there is no twofold rotation around the in-plane direction that leave both of them invariant. This state is symmetric only under $\mathcal{P}$ and $\mathcal{T}C_{2z}$ and its MSG is P2$'$/m$'$, as for the FIM$ab$ and FF$ab$. 

The third category is the canted ($\pm$) states. In these states, the Mn and $R$ moments have unrelated orientations within a general plane whose normal points to a tilted direction in space. None of the rotation operations, even if combined with time-reversal are preserved in these states. The only symmetry left in the canted ($\pm$) states is a spatial inversion and its MSG is P$\bar{1}$. 

{\it Spiral states}. 
Another category of coplanar states corresponds to the triple-spiral states. In this state, each Mn and $R$ sublattice forms a simple helical spiral along $\hat{z}$ with the same period and an arbitrary phase rotation between each sublattice. This state is similar to the HP($\pm$) in the sense that the Mn and $R$ moments lie along noncollinear directions within the $ab$-plane.
However, the triple spiral state has nonzero vector spin chirality where inversion symmetry is broken since the Mn moment directions in the inversion-related Mn layers are different. Therefore, the triple-spiral state is invariant only under $\mathcal{T}C_{2z}$ and its MSG corresponds to P2$'$ (No. 3.3).

The least symmetrical state we encounter is the vertical conical spiral (VCS).  Here, each Mn and $R$ sublattice moment that forms a helical triple-spiral is also canted along the $\hat{z}$ axis. This state also has vector spin chirality where all symmetries of the parent gray group are broken and its MSG has only the identity operation, group P1 (No. 1.1). Note that $\mathcal{P}$ is preserved in all magnetic states considered except for the triple-spiral and VCS phase (see Table \ref{tbl:phases}).

A summary of the MSGs of the possible ground states of our model as well as illustrations of representatives of these families of magnetic states can be found in Table \ref{tbl:phases}. Next, we provide a detailed discussion of the phases that can be reached in each member of the $R$166 family and describe the nature of the phase transitions.

\section{Calculation details}

 Employing the comprehensive mathematical framework and equations previously discussed, we compute the free energy as a function of the magnetic moment orientation angles.  For a given temperature and applied field and set of orientation angles, we self-consistently determine the spin magnitudes that produce the correct molecular field.  With this knowledge, we minimize the free energy using a search procedure that is seeded by random values of the orientation angles. The moment angles and average sublattice magnetizations for the minimum-energy configuration were analyzed to determine the magnetic phase and symmetry according to Table~\ref{tbl:phases}. These calculations enabled the construction of the final phase diagrams for each material as a function of temperature and magnetic field.

\begin{figure}[ht]
\includegraphics[width=0.95\linewidth,clip]{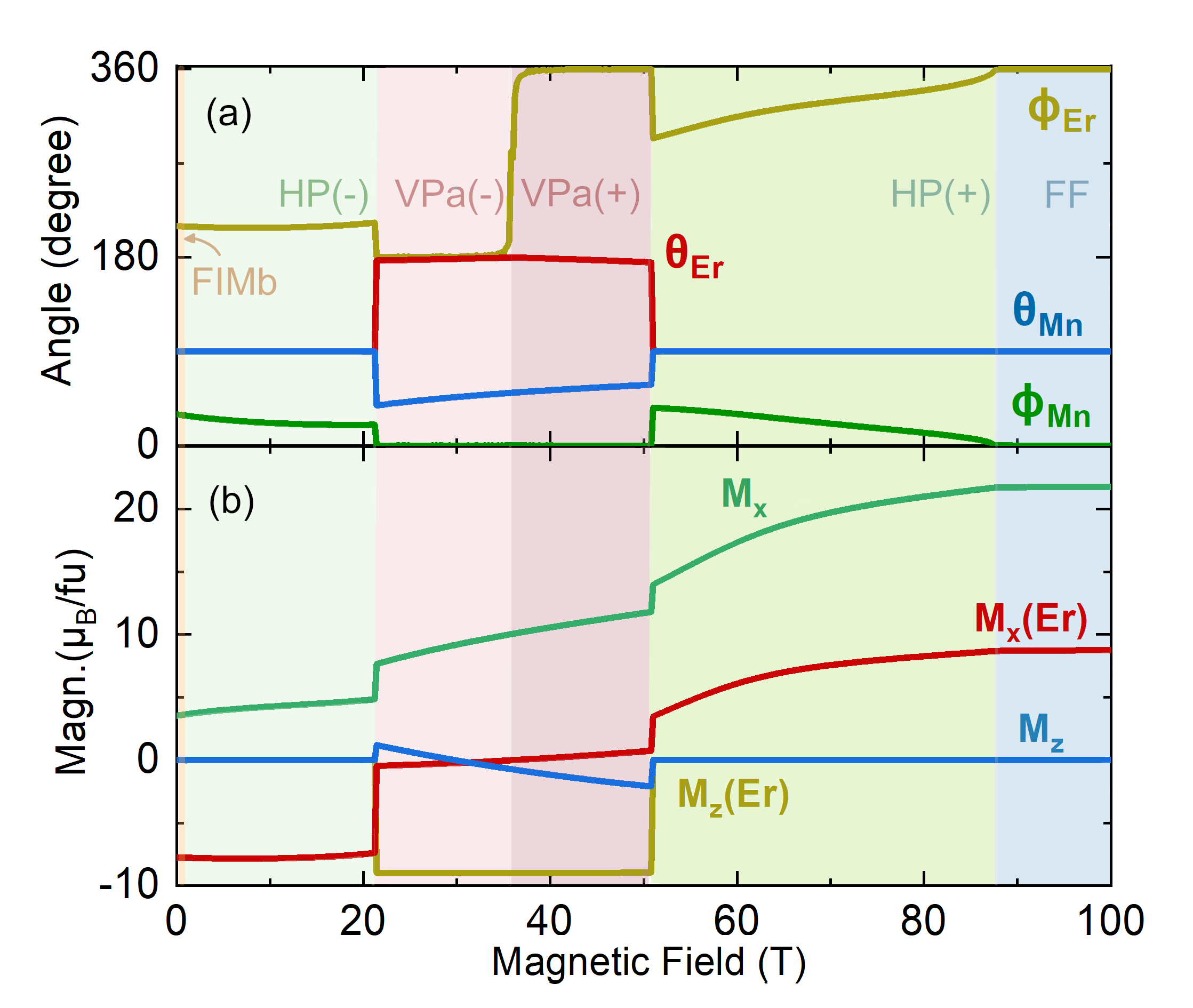} 
\caption{Evolution of magnetic phases in ErMn$_6$Sn$_6$ under an applied magnetic field parallel to the $x$-axis (${\bf H}\parallel a$) at $T=0$ K. (a) Spherical angles of each sublattice and (b) Er magnetization components and the total magnetization along the $x$ and $z$ directions as functions of the magnetic field.  Different phases are labeled using the notation described in Table~\ref{tbl:phases}. }
\label{fig:Er_a_compare}
\end{figure}

Selected magnetization calculations and other results can be found in Appendix B that may be readily compared with published data. Taking Er166 with an applied magnetic field along the $x$-axis (${\bf H}\parallel a$) as an example, the evolution of the moment angles and magnetization curves provide critical insights into phase transitions as shown in Fig.~\ref{fig:Er_a_compare}. For example, the low-field results below 20 T indicate that both Er and Mn moments lie in the $ab$ plane ($\theta^{E} = \theta^{M}=90 \degree$).  In the FIM$b$ phase at zero-field, the Er and Mn moments point along the planar easy-axis at $\varphi^M =30 \degree$ and $\varphi^E = \varphi^M+180\degree$.  A small field with ${\bf H}\parallel a$ distorts the FIM$b$ phase into the horizontal plane canted (HP) phase with smoothly evolving planar angles where $\varphi^E-\varphi^M \neq 180 \degree$ and smoothly increasing magnetization.

Sharp discontinuities or abrupt changes in the magnetization ($M_{\alpha}=\partial \mathcal{F}/\partial H_{\alpha}$) signify first-order transitions to another magnetic phase. For example, in the first-order transition from HP($-$) to a vertical plane canted structure [VP$a$($-$)] near 20 T, all magnetization components are discontinuous as moments jump out of the layer. In contrast, discontinuities in the slope of the magnetization ($\partial^2 \mathcal{F}/\partial H^2$), such as from HP(+) to FF, indicate second-order transitions. Smooth curves or gradual variations typically imply no transition or a crossover unless a symmetry change occurs. For example, the crossover from VP$a$($-$) to VP$a$($+$), where the Er sublattice magnetization crosses $M_x^{Er}=0$, is marked even though there is no real transition. 

To address more subtle variations with weak thermodynamic signatures or vertical phase lines in the $T-H$ phase diagrams, we sometimes used constant-field temperature sweeps.  Here, we can compute the entropy (-$\partial \mathcal{F}/\partial T$) and the heat capacity as a secondary reference and increase precision in certain regions to optimize prediction accuracy. Despite these efforts, minor point-by-point errors in the assignment of phase lines are unavoidable, although they were minimized as much as possible.

Phase diagrams are assessed in this manner for each material to identify phase transitions across temperatures up to 300 K or until all moments align with the applied field. We note that our mean-field calculations overestimate the magnetic ordering temperature ($T_C$) that is driven by large intralayer Mn-Mn exchange. Experimental results across the series find $T_C \approx$ 370-430 K whereas our mean-field calculations find $T_C \approx$ 1000 K.  This overestimation is typical for mean-field analysis of layered systems and affects the accuracy of our mean-field calculations for $T\gtrsim 250 K$.

\section{Results}

 \subsection{GdMn$_6$Sn$_6$}
 \begin{figure*}[ht]
\centering
    \includegraphics[width=1\linewidth]{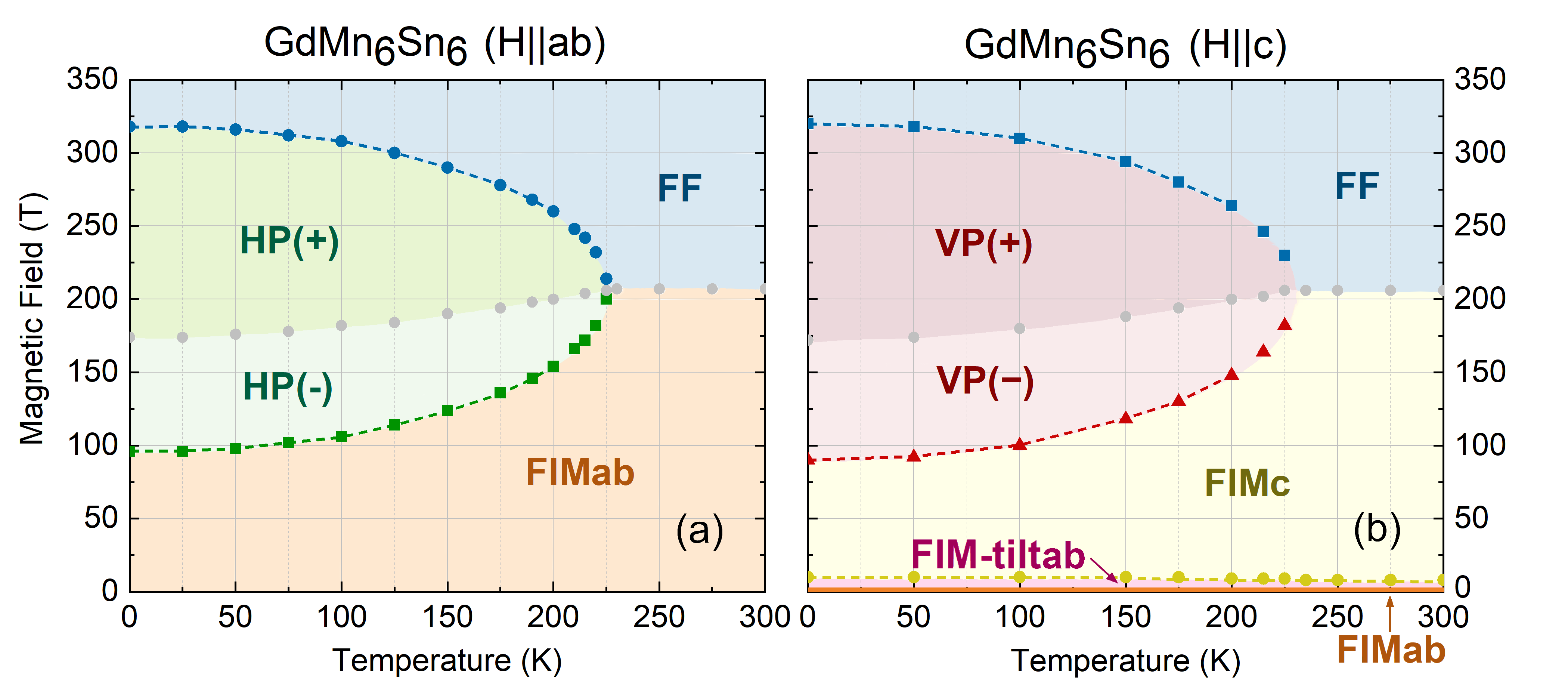}
\caption{Mean-field magnetic phase diagrams for GdMn$_6$Sn$_6$ for fields in (a) the $ab$-plane and (c) along the $c$ axis.  Labels for the different magnetic phases are defined in Table \ref{tbl:phases}. Colored symbols correspond to phase transitions determined from mean-field calculations. Dashed lines indicate second-order-like phase transitions.  Gray circles mark the crossover lines corresponding to $H^R$ and $H_{\parallel}^R(T)$.}
\label{fig:Gd}
\end{figure*}
 
Gd166 is the simplest of the $R$166 compounds since Gd is a spin-only ion with no magnetic anisotropy (all $B_l^m=0$) and only an easy-plane Mn anisotropy is present.  The AF Gd-Mn exchange coupling is the largest of all $R$166 compounds \cite{Tils1998}, therefore the zero-field structure is a planar (easy-plane) ferrimagnet (FIM$ab$) \cite{Venturini1991}. 
 
Fig.~\ref{fig:Gd} shows the phase diagrams for ${\bf H} \parallel ab$ and ${\bf H} \parallel c$ in the coplanar model.  Both phase diagrams are very similar with small differences caused by the easy-plane anisotropy of the Mn ion. For example, a small applied field along $c$ at low temperatures will coherently rotate the saturated net ferrimagnetic magnetization, $M=(6gs-g_RJ)\mu_B = 6~\mu_B$, through the FIM-tilt$ab$ phase and towards the $c$-axis. An estimated critical field of $\mu_0H=12D^Ms^2/M=9$ T is required to overcome the planar Mn anisotropy and line up the ferrimagnetic moment with the field (FIM$c$, as shown in Fig.~\ref{fig:R166_mag}(a) in Appendix B). This is consistent with magnetization data that report a critical field of 9-10 T \cite{Gorbunov2012, Kimura2006}.

However, both experimental magnetization measurements disagree with each other at high field strengths.  In Ref.~\cite{Kimura2006}, the experimental magnetization does not fully saturate at 10 T and has a weak linear dependence up to 55 T where $M\approx 6~\mu_B$ is close to saturation. In Ref.~\cite{Gorbunov2012}, the experimental magnetization is $\approx6~\mu_B$ at 10 T and grows to $10~\mu_B$ at 60 T.  The origin of the discrepancy between these two measurements not known, however, the latter result would indicate some degree of canting already at small fields. In comparison, our mean-field calculations predict that the $6~\mu_B$ magnetization plateau remains until the canted phase is entered near 91 T in better agreement with Ref.~\cite{Kimura2006}.

For either field direction, much larger field strengths of order $\sim$300 T are required to overcome $\mathcal{J}^{MR}$ and align both sublattices parallel to the field (FF phase). This occurs through a canted phase which is entered and exited via second-order transitions.  For  ${\bf H} \parallel c$, the transverse magnetization component in the VP phase has $XY$ symmetry due to the easy-plane anisotropy. 
 
Given the simplicity of Gd166, we can compare our mean-field results to analytical calculations to confirm the accuracy of our approach. At $T=0$, the critical fields for these transitions (in the absence of single-ion anisotropy) are given by (see Appendix A for derivation)
 \begin{equation}
     \mu_0H_c^{\pm} = \frac{2\mathcal{J}^{MR}(g_R-1)}{gg_R\mu_B}[6g s \pm g_R J].
     \label{eqn:Hc}
 \end{equation}
The formula above gives $\mu_0H_c^-=96$ T and $\mu_0H_c^+=318$ T in agreement with mean-field calculations.

\begin{table}
\caption{Characteristic fields for $R$Mn$_6$Sn$_6$ compounds (in Tesla) calculated from Eqs. \ref{eqn:comp_field}, \ref{eqn:Hc}, and \ref{eqn:HR0} for an isotropic system.}
\renewcommand\arraystretch{1.25}
\centering
\begin{tabular}{ c | c | c | c | c }
\hline\hline
$R$  	& $\mu_0 H_c^-$  & $\mu_0 H_{\parallel}^R(0)$	& $\mu_0 H^R$ 	& $\mu_0 H_c^+$   \\
\hline					
Gd &		96	& 174	& 207    & 318		  \\
Tb &		39	& 91	& 126    & 213		  \\
Dy &		20	& 55	& 86    & 153		  \\
Ho &		14	& 40	& 62    & 110		  \\
Er &		14	& 34	& 47    & 79		  \\
Tm &		16	& 30	& 35    & 54		  \\
\hline\hline
\end{tabular}
\label{tbl:fields}
\end{table}
 
As the temperature increases, thermal fluctuations will reduce the ordered Gd moment $J\rightarrow\langle J\rangle$ while the Mn magnetic sublattice remains rigid ($\langle s\rangle = s$) due to the large Mn-Mn intralayer exchange ($\mathcal{J}_0^{MM}$). According to Eq.~\ref{eqn:Hc}, this leads to an increase (decrease) in $H_c^-$ ($H_c^+$), respectively, (see Appendix A for details) and the two transitions merge when $\langle J\rangle\approx 0$.  This occurs at an applied field (called the compensation field) of 
 \begin{equation}
 \mu_0H^R = \frac{12\mathcal{J}^{MR}(g_R-1)}{g_R\mu_B}\langle s\rangle = 207~{\rm T}
 \label{eqn:comp_field}
 \end{equation}
 which is the strength necessary to cancel the molecular field acting on the $R$ site (see Eq.~\ref{eqn:BR_co}).  
 
 At the compensation field $H^R$, we define a critical temperature labeled $T_c^R$. Above $T_c^R$, the Gd moment is completely quenched and has no preferred direction at $H^R$ ({$\langle {\bf J}\rangle\approx 0$). The field evolution from FIM$ab$ or FIM$c$ to FF occurs without an intervening canted phase when $T>T_c^R$.  Rather, there is a crossover where the Gd moment reverses direction but always lies parallel or antiparallel to the applied field. 
 
 In the canted phase below $T_c^R$, the compensation field can be extended to define a line [$H_{\parallel}^R(T)$] indicating the crossover between canted ($-$) and canted ($+$) phases. Following this line, the $R$ moment parallel to the applied field vanishes ($\langle J_{\parallel}\rangle = 0$) as this component is compensated by the applied field. Thus, the $R$ ion is only acted on by the perpendicular molecular-field component caused by canting and this component will order below $T_c^R$ $(\langle J_{\perp}\rangle \neq 0$).
 
We can calculate the value of the parallel compensation field $H_{\parallel}^R$ for Gd166 at $T=0$. For a fully isotropic model, the net moment perpendicular to the applied field is always zero ($M_\perp = 0$). We use this to find the maximum Mn canting angle of $\sin\varphi^M_ {max}=g_RJ/6gs$ which occurs when the full Gd moment lies perpendicular to the applied field.  This occurs at an applied field strength of  
\begin{equation}
    \mu_0H_{\parallel}^R(0)=\mu_0H^R\cos\varphi^M_ {max} = \mu_0H^R\sqrt{1-\Big(\frac{g_RJ}{6gs}\Big)^2}.
    \label{eqn:HR0}
\end{equation}
For Gd166, we find that $\mu_0H^R(0)=174$ T in agreement with mean-field results.  A numerical derivation of the full $H^R_{\parallel}(T)$ line is provided in Appendix A.

When cooling Gd166 at a field strength of $H_{\parallel}^R(T)$, $T_c^R=227$ K is the critical temperature below which the Gd moment perpendicular to the field orders.  We can compare this situation to the ordering of a simple single-sublattice ferromagnet shown in Fig.~\ref{fig:Gd_effective}. For the ferromagnet, cooling below the critical temperature $T_c$ will result in the coexistence of up and down magnetic domains on a first-order line at $H=0$. We compare this to the response of the Gd sublattice in an effective field with components given by
\begin{equation}
H'_{||} = H - H^R \cos\varphi^M
\end{equation}
and
\begin{equation}
H'_{\perp} =  - H^R \sin\varphi^M.
\end{equation}
Here, $\varphi^M$ is the canting angle of the Mn sublattice away from the field direction, $H_{\parallel}'$ is the applied field reduced by the parallel molecular field component, and $H_{\perp}'$ is the perpendicular component of the molecular field. Along $H_{\parallel}^R(T)$, $H_{\parallel}^{\prime} = 0$ and  $H_{\perp}'\neq0$ below the critical point. Thus for Gd166, the first-order line at $H_{\perp}'=0$ is not physically accessible and the system follows the dashed line upon cooling.

\begin{figure}[ht]
\includegraphics[width=0.9\linewidth,clip]{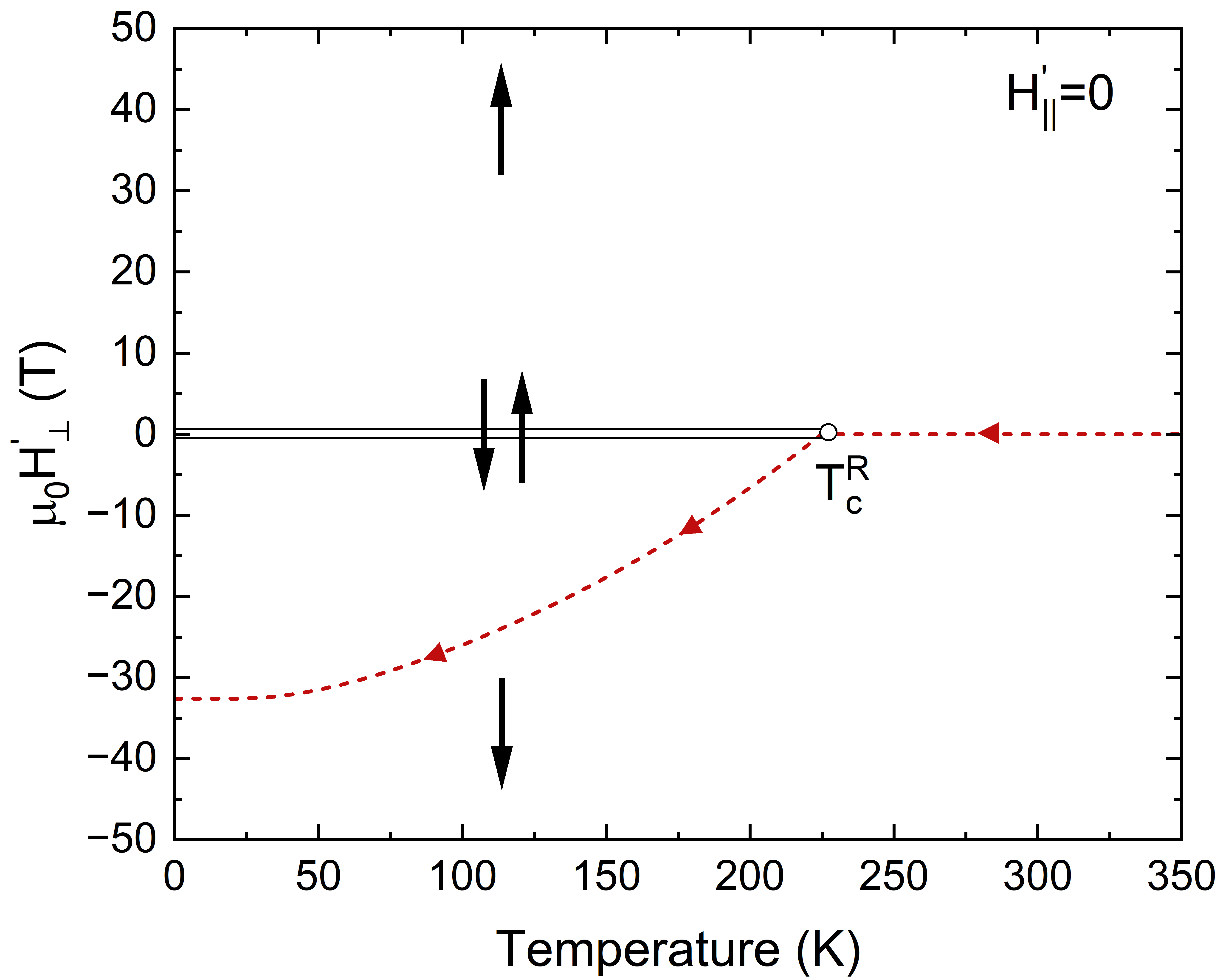} 
\caption{ The phase diagram of a simple ferromagnet where cooling through the critical point $T_c^R$ results in a first-order line (double line) with coexistence of up and down magnetic domains. Lowering the temperature for a nonzero $H_{\perp}^{\prime}$ field results in the crossover to a field polarized state.  For GdMn$_6$Sn$_6$ with $H_{\parallel}^{\prime}=0$, the first-order line is not physically accessible as the canting of the Mn sublattice moment forces the Gd sublattice to follow the red-dashed path in the $H_{\perp}^{\prime}-T$ plane (along the $H_{\parallel}^R(T)$ line).
}
\label{fig:Gd_effective}
\end{figure}

\subsection{TbMn$_6$Sn$_6$}

\begin{figure*}[ht]
\includegraphics[width=1\linewidth,clip]{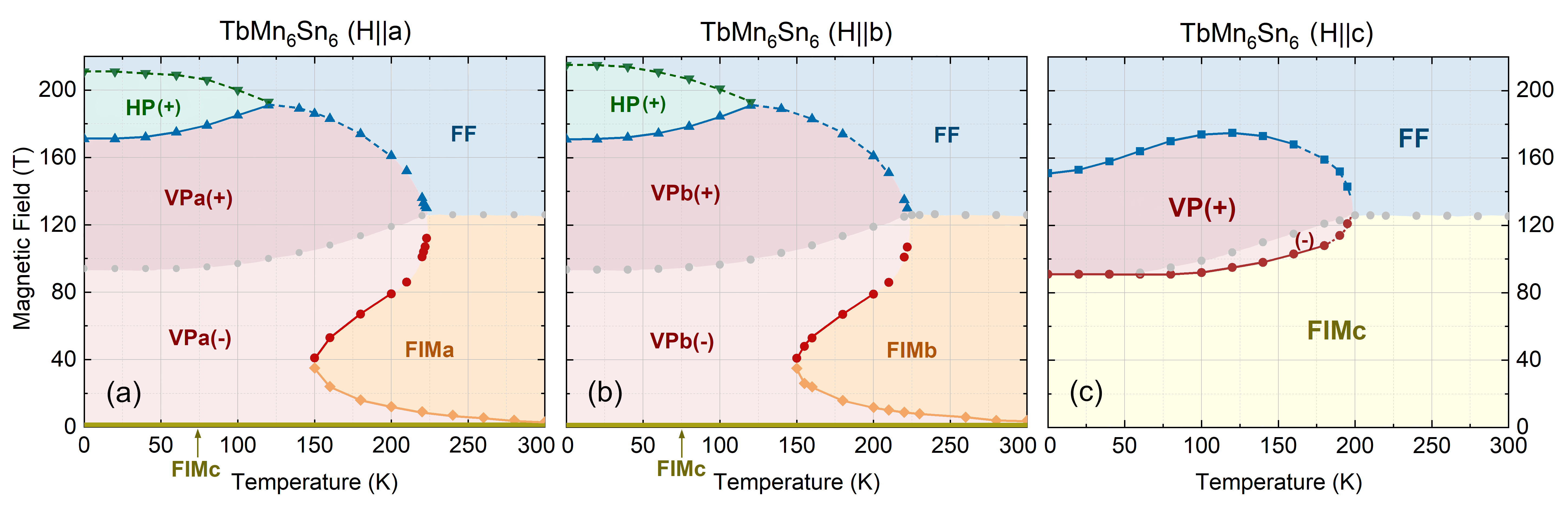} 
\caption{Mean-field magnetic phase diagrams for TbMn$_6$Sn$_6$ for fields along $a$, $b$, and $c$ directions.  Labels for the different magnetic phases are defined in Table \ref{tbl:phases}.  Colored symbols correspond to phase transitions determined from mean-field calculations. Solid (dashed) lines correspond to first-order-like (second-order-like) phase transitions, respectively.  Gray circles mark the crossover lines corresponding to $H^R$ and $H_{\parallel}^R(T)$.}
\label{fig:Tb}
\end{figure*}

 Tb166 is characterized by a large uniaxial Tb anisotropy which competes with the easy-plane Mn anisotropy. Combined with large AF Tb-Mn exchange coupling \cite{Riberolles2022}, the zero-field structure is FIM$c$ at low temperatures. At high temperatures, the Tb anisotropy is sufficiently quenched by thermal fluctuations \cite{Riberolles2023} that the easy-plane Mn anisotropy becomes dominant and the system undergoes a first-order spin-reorientation transition into the FIM$a$ phase (the $a$-axis is the planar easy-axis for $B_6^6<0$).  The spin-reorientation transition is observed at $T_{SR}\approx 310$ K \cite{Clatterbuck1999} while our mean-field calculations give $T_{SR}= 367$ K (see Fig.~\ref{fig:R166_SR} in Appendix B) which suggests that our CEF parameters overestimate the uniaxial anisotropy for Tb166.
 
 Figure~\ref{fig:Tb}(c) shows the phase diagram for ${\bf H} \parallel c$ in the coplanar model. The FIM$c$ phase is stable up to very high fields of order 100 T. The nominal critical fields that bound the VP$a$ phase in the absence of anisotropy (Eq.~\ref{eqn:Hc} and Table~\ref{tbl:fields}) at low temperatures are $\mu_0H_c^{-}=39$ T and $\mu_0H_c^{+} = 214$ T.  However, the stability regime of the VP$a$ phase is strongly reduced in Tb166 by the Tb uniaxial anisotropy. The entry into the VP$a$($-$) phase occurs through a first-order spin-flop transition at 91 T where Tb moment flops into the basal plane (Tb-flop) while the Mn moment cants away from the $c$-axis ($\theta^M_{max} = 44 \degree$) to minimize the $\mathcal{J}^{MR}$ exchange coupling energy. Here, the spin-flop transition coincides with the $H_{\parallel}^R(T)$ line at low temperatures. The first-order nature of the spin-flop is caused by the high-order anisotropy term ($B_4^0$ and $K_2$, see Table~\ref{tbl:CEF}) that creates a large energy barrier between the uniaxial and planar Tb orientations. Another first-order transition occurs upon exiting the VP$a$($+$) phase and into the FF phase at 153 T (Tb-flip). Here the Tb and Mn moment both jump parallel to the field leading to full saturation of the magnetization.

At elevated temperatures, the two critical-field lines evolve from first-order-like to second-order-like (a tricritical point) due to thermal fluctuations that weaken the Tb anisotropy. The VP$a$ phase is eventually squeezed out above a critical temperature of $T_c^R\approx 200$ K at the Tb compensation field of $\mu_0H^R=126$ T. Similar to Gd166, the Tb moment is completely quenched at $H^R$ and above $T_c^R$. Re-entrant behavior of the FF phase is observed when cooling at fields above $H^R$ and to a lesser extent below $H^R$. The re-entrant phenomenon is caused by weakening anisotropy initially increasing the stability range of the canted phase before it disappears near $T_c^R$.

We now consider the Tb166 phase diagram for in-plane fields with ${\bf H}\parallel a$ and ${\bf H}\parallel b$ shown in Fig.~\ref{fig:Tb}.  These are similar to each other, and we discuss only the ${\bf H}\parallel a$ case where the field direction coincides with the planar easy-axis. Starting at the low-temperature and zero-field FIM$c$ phase, the Mn moment cants towards the field. The Tb moment initially cants away from the field [VP$a$($-$)] to minimize the exchange before reversing towards the field into the VP$a$($+$) phase at 94 T (close to $\mu_0H_{\parallel}^R(0)$ and similar to the spin-flop field for ${\bf H}\parallel c$). At even higher fields, the large Tb anisotropy energy barrier for $\theta^R$ rotation triggers a jump of the Tb and Mn moments into the basal plane. This first-order transition from VP$a$($+$) to HP($+$) phase occurs near 170 T. Finally, the Tb-Mn angle in the HP($+$) phase continuously closes to the FF phase near $\mu_0H_c^{+} = 214$ T.

Similar to the ${\bf H}\parallel c$ phase diagram, higher temperature evolution of the ${\bf H}\parallel a$ or $b$ phases is controlled by thermal quenching of the Tb anisotropy and reduction of the Tb moment. Ultimately, collinear, easy-plane phases are favored at high temperatures. The HP phase is squeezed out near (120 K, 192 T) at a bicritical point where the HP$-$VP$a$ coexistence line intersects with the second-order FF phase boundary. The VP$a$ phase is also squeezed out at $(T_c^R,\mu_0H^R) \approx$ (224 K, 126 T). 

Additional complexity in the ${\bf H}\parallel a$ phase diagram at high temperatures and low fields is caused by appearance of the FIM$a$ phase. At zero-field, we find a first-order spin reorientation transition from FIM$c$ to FIM$a$ occurs at $T_{SR}=367$ K. Below $T_{SR}$, a weak in-plane field along the hard $a$-axis triggers a first-order metamagnetic transition (FOMP) from FIM$c$ (VP$a$) phase.  Our mean-field calculations predict a critical FOMP field of 8.9 T at 200 K that is slightly larger than the critical field of 7.8 T measured on single-crystal samples at the same temperature \cite{Clatterbuck1999}. Continuing to increase the field along the magnetization direction of FIM$a$ for $T<T_c^R$ results in a first-order spin-flop transition that re-enters the VP$a$ phase. As temperature is lowered, increasing uniaxial anisotropy both increases the FOMP field and lowers the spin-flop field causing the transitions to merge at (150 K,39 T).

\subsection{DyMn$_6$Sn$_6$ and HoMn$_6$Sn$_6$}
\begin{figure*}[ht]
\includegraphics[width=1\linewidth,clip]{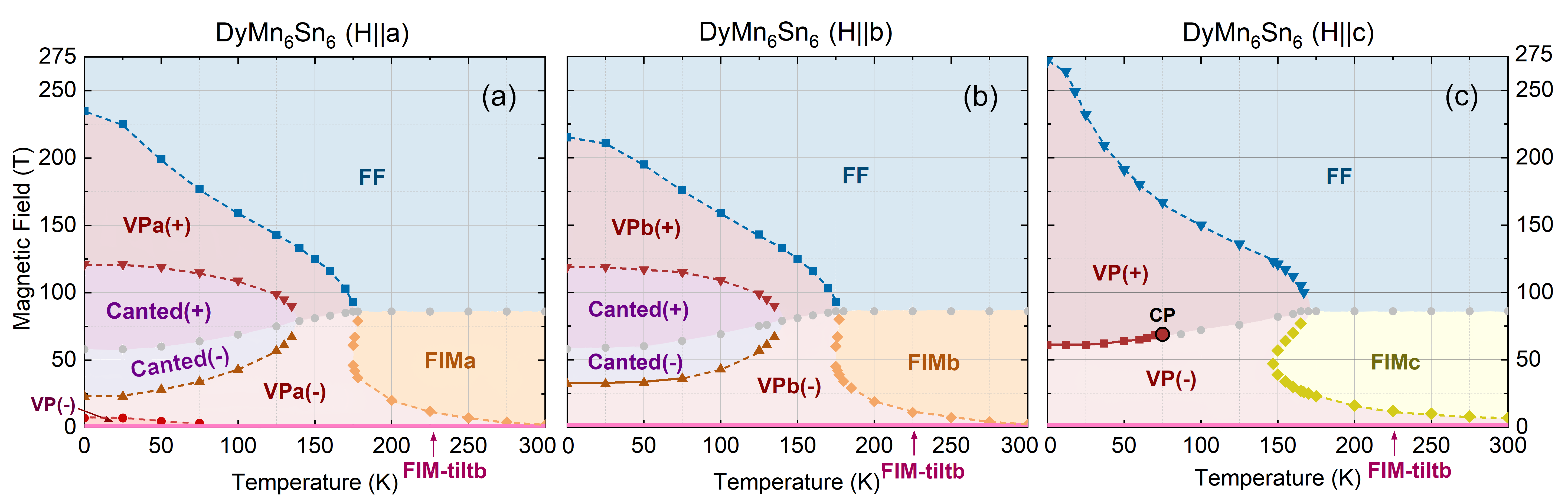}
\caption{Mean-field magnetic phase diagrams for DyMn$_6$Sn$_6$ for fields along (a) $a$, (b) $b$, and (c) $c$ directions. Labels for the different magnetic phases are defined in Table \ref{tbl:phases}. Colored symbols correspond to phase transitions determined from mean-field calculations. Solid (dashed) lines correspond to first-order-like (second-order-like) phase transitions, respectively.  Gray circles mark the crossover lines corresponding to $H^R$ and $H_{\parallel}^R(T)$. CP labels a liquid-gas-like critical point.
}
\label{fig:Dy}
\end{figure*}

\begin{figure*}[ht]
\includegraphics[width=1\linewidth]{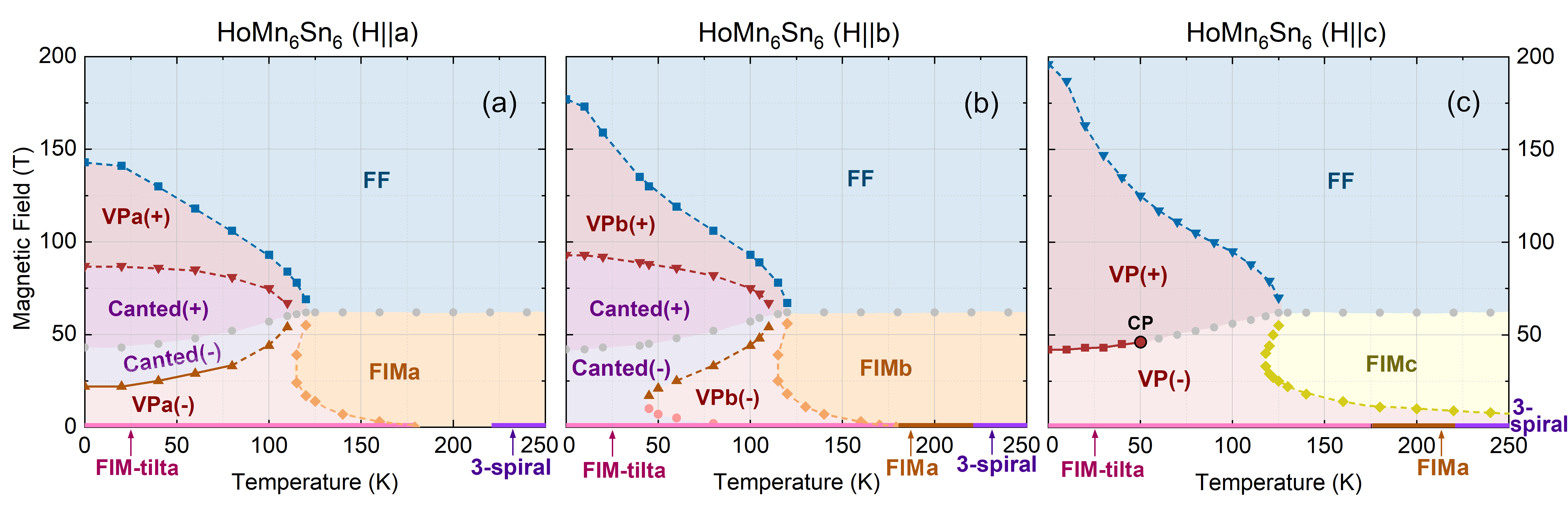}
\caption{Mean-field magnetic phase diagrams for HoMn$_6$Sn$_6$ for fields along (a) $a$, (b) $b$, and (c) $c$ directions. Labels for the different magnetic phases are defined in Table \ref{tbl:phases}.  Colored symbols correspond to phase transitions determined from mean-field calculations. Solid (dashed) lines correspond to first-order-like (second-order-like) phase transitions, respectively.  Gray circles mark the crossover lines corresponding to $H^R$ and $H_{\parallel}^R(T)$.  CP labels a liquid-gas-like critical point.}
\label{fig:Ho}
\end{figure*}

Both DyMn$_6$Sn$_6$ and HoMn$_6$Sn$_6$ possess relatively large $K_1$ and $K_2$ MAE constants of opposite sign (see Table \ref{tbl:CEF}).  Unlike Tb, however, $K_1<0$ and $K_2>0$ resulting in an easy-cone geometry where Dy or Ho moments prefer to tilt away from the $c$-axis. Thus, Dy166 and Ho166 nominally adopt a FIM-tilt structure with calculated $R$ tilt angles of approximately $46\degree$ and $48\degree$ at low temperatures and zero field, respectively. These agree with neutron-diffraction analysis reporting tilt angles of $45\degree$ \cite{Malaman1999} and $49\degree$ \cite{ElIdrissi1991}.  Due to the finite $\mathcal{J}^{MR}$ coupling and Mn easy-plane anisotropy, our calculations indicate that} $R$ and Mn moments are not exactly collinear in the ground state and should formally be considered to reside in a VP-canted phase rather than a FIM-tilt phase. These two phases have the same magnetic space group.

Here we describe the specific case of Dy166 shown in Fig.~\ref{fig:Dy} while keeping in mind that the Dy166 and Ho166 phase diagrams (shown in Fig.~\ref{fig:Ho}) are qualitatively similar. For Dy166, the planar easy-axis is along $b$ and it adopts the VP$b$ (FIM-tilt$b$) phase in zero field. At low temperatures, the application ${\bf H} \parallel a$ causes a coherent rotation of the net magnetization from the VP$b$ phase to the VP$a$ phase at 6 T. Near 23 T, the $R$ moment begins to rotate around the $c$-axis (the easy-cone direction) causing $\varphi^R$ to rotate towards the field. Minimization of the exchange leads to a nonproportional rotation of the Mn moment ($\varphi^M$) causing the canting plane to tilt in a general direction. Note that, for ${\bf H} \parallel b$ (the planar easy-axis), this transition occurs at a critical field of 31 T and is first-order, corresponding to a jump in the Dy planar angle $\Delta\varphi$ similar to a spin-flop transition.

One difference between Ho166 and Dy166 phase diagrams can be seen in Fig.~\ref{fig:Ho}(b) for Ho166 with ${\bf H}\parallel b$. Ho166 is VP$a$ (FIM-tilt$a$) at zero-field and has a significantly larger in-plane anisotropy constant $K_3'$ than Dy166 (see Table \ref{tbl:CEF}). As a consequence, the Ho166 does not rotate coherently into the VP$b$ phase at low temperatures. Instead, it enters a general canted phase. At temperatures above 40 K, the in-plane anisotropy is thermally softened enough to allow the VP$b$ phase to appear.

These phase diagram features are consistent with high-field magnetization measurements of the critical field of the VP-to-canted transition for both Dy166 and Ho166 \cite{Kimura2006}. For Dy166, a weak second-order transition starting near $\approx ~25$ T and a first-order transition at 31 T are observed for ${\bf H}\parallel a$ and ${\bf H}\parallel b$, respectively.  We compare this to our mean-field calculations that find a second-order transition with a weak magnetization signature at 23 T and a first-order transition at 31 T. For Ho166, a first-order transition at $\approx ~22$ T and a 26 T feature with a weak magnetization signature is observed for ${\bf H}\parallel a$ and ${\bf H}\parallel b$, respectively. Analysis of our mean-field calculations for Ho166 largely agree with these observations.  We find a first-order transition at 23 T for ${\bf H}\parallel a$ and a crossover with a broad magnetization signature in the range of 20 to 30 T when ${\bf H}\parallel b$. This crossover occurs due to the backbending of the VPb-canted phase line shown in Fig.~\ref{fig:Ho}(b). Figures \ref{fig:R166_mag}(c) and (d) in Appendix B show these magnetization plots.

At even higher planar fields, the component of the $R$ moment parallel to the field passes through zero at $\mu_0H^R_{\parallel}(0)=55$ T and returns to VP$a$ phase near 119 T.  Above this critical field, the angle between Dy and Mn moments closes towards the field, becoming fully aligned into the FF phase at 235 T. These critical fields can be compared to the characteristic fields for an isotropic system in Table~\ref{tbl:fields}.

At elevated temperatures above $T^R_c\approx 175$ K, quenching of the Dy anisotropy squeezes out canted phases in favor of collinear FIM$a$ or FF phases. At zero-field, the quenching leads to the a spin-reorientation transition into the FIM$a$ phase at $T_{SR}$. Unlike Tb166, the spin-reorientation transitions in Dy166 and Ho166 are continuous. The experimental critical temperatures of $T_{SR}\approx300$ K and $\approx 200$ K for Dy166 and Ho166 \cite{Malaman1999}, respectively, compare favorably to our mean-field calculations of 328 K and 175 K, as shown in Appendix B Fig.~\ref{fig:R166_SR}(a).  The SR transition is continuous for Dy166 and Ho166 since no intermediate energy barrier for polar rotation exists with easy-cone anisotropy. A field-driven transition into the FIM$a$ phase for $T<T_{SR}$ occurs through rotation of the net magnetization and is also second-order.

Whereas the phase diagram for ${\bf H}\parallel b$ is similar to ${\bf H}\parallel a$, the ${\bf H}\parallel c$ phase diagram for both Dy166 and Ho166 has an interesting feature. Again, we refer to the Dy166 phase diagram shown in Fig.~\ref{fig:Dy}(c).  At low temperatures, the field ${\bf H}\parallel c$ leads to a strong first-order transition at 62 T corresponding to the coexistence of VP$b$($-$) and VP$b$($+$) phases.  The same transition is predicted to occur at 45 T for Ho166 [see Appendix B Fig.~\ref{fig:R166_SR}(a)] and has been observed in high-field measurements at this field strength \cite{Kimura2006}. The first-order nature occurs because polar angle of the Dy (Ho) moment is either in the MAE minimum where $M_R^z$ is antiparallel to the field ($\approx133\degree$) or $M_R^z$ is parallel to the field ($\approx47\degree$). These states have the same magnetic symmetry and $\Delta M_R^z$ acts as (discontinuous) order parameter. As temperature is increased, quenching of the anisotropy rotates the Dy moment towards the $ab$-plane. This shrinks the discontinuity $\Delta M_R^z$ which reaches zero at a liquid-gas-like critical point (CP). Beyond the critical point, only a crossover from VP$b$($-$) to VP$b$($+$) occurs.  At temperatures above $\approx 150$ K, a small field generates a FOMP-like transition to the FIM$c$ phase although, unlike Tb166, this transition is second-order due to the easy-cone anisotropy. The FIM$b$ phase crosses over to the FF phase at $\mu_0H^R=86$ T.

\subsection{ErMn$_6$Sn$_6$}

 \begin{figure*}[ht]
\centering
\begin{tabular}{ccc}
  \includegraphics[width=1\linewidth,clip]{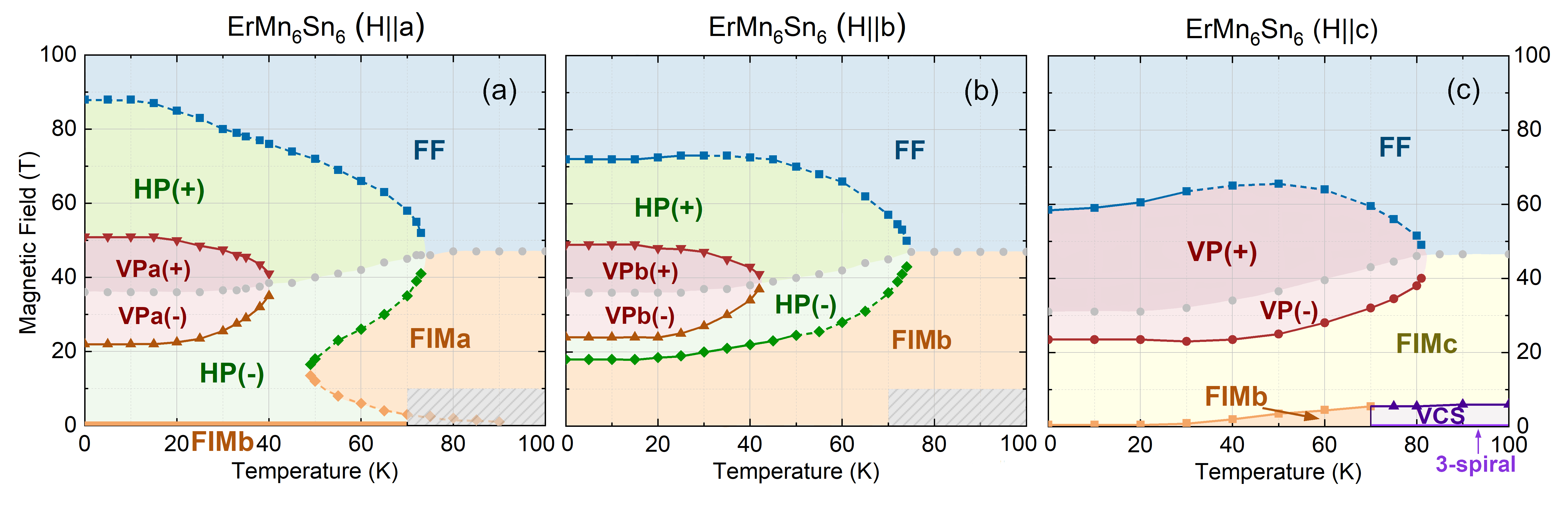}
\end{tabular}%
\caption{Mean-field magnetic phase diagrams for ErMn$_6$Sn$_6$ for fields along (a) $a$, (b) $b$, and (c) $c$ directions.  Labels for the different magnetic phases are defined in Table \ref{tbl:phases}.  Colored symbols correspond to phase transitions determined from mean-field calculations. Solid (dashed) lines correspond to first-order-like (second-order-like) phase transitions, respectively.  Gray circles mark the crossover lines corresponding to $H^R$ and $H_{\parallel}^R(T)$. Hatched areas correspond to regions where noncollinear conical and fan-like phases appear that are outside our analysis method.  Transitions and critical points are labeled.}
\label{fig:Er}
\end{figure*}
Er166 is perhaps most complex member of the $R$166 series. Magnetization data \cite{Suga2006} and mean-field results \cite{Riberolles2024b} confirm that the zero-field ground state is FIM$b$. However, the weak Er polar MAE leads to near degeneracy between FIM$b$ and FIM$c$ which results in a small critical field ($< 1$ T) for the FOMP transition with ${\bf H}\parallel c$ \cite{Clatterbuck1999}. Furthermore, the presence of weaker $\mathcal{J}^{MR}$ exchange drives the system into a noncollinear spiral phase with nonzero vector chirality at temperatures above $T_{spiral}\approx 70$ K \cite{Malaman1999, Riberolles2024b}.  Essentially, the FIM$b$, FIM$c$, and triple-spiral order are all nearly degenerate in zero-field.

We discuss the phases in applied field by starting with ${\bf H}\parallel c$ in Fig.~\ref{fig:Er}(c).  At low temperatures, fields above a critical strength of 0.8 T will stabilize FIM$c$ via a FOMP transition. Similar to Gd166 and Tb166, Er166 enters the VP$b$($-$) phase via a first-order spin-flop transition at 25 T where the Er moment flops into the plane, causing the Mn moment to cant away from the $c$-axis (Er-flop). This spin-flop transition has been observed also at 25 T in high-field magnetization measurements \cite{Suga2006}. While the Mn moment cants back towards the field in the VP$b$ phase at higher fields, the Er moment only weakly rotates towards the field until it flips parallel to the field in a first-order transition at 58 T (Er-flip) and enters the FF phase. Current experimental data do not exceed 50 T, so this transition has not yet been observed. As temperature increases, the field range over which the VP$b$ phase is stable first increases, similar to Tb. Both the Er-flop and Er-flip transition lines become second-order-like before the VP$b$ phase is squeezed out near $T_c^R=85$ K and $\mu_0H^R=47$ T.

At fields below 10 T, the competition between FIM$b$, FIM$c$, and triple-spiral order plays out as the temperature increases. Our mean-field calculations ultimately favor the triple-spiral phase above $T_{spiral} = 75$ K with a strongly temperature dependent pitch angle [see Appendix B Fig.~\ref{fig:R166_SR}(b)].  These results are in excellent agreement with neutron-scattering data \cite{Riberolles2024b}. Starting at the triple-spiral phase, ${\bf H}\parallel c$ stabilizes the vertical conical spiral (VCS) where moments cant out of the plane. At temperatures below $T_{spiral}$, the FIM$b$-to-FIM$c$ FOMP transition is first-order.  Above $T_{spiral}$, the VCS-to-FIM$c$ transition is initially a first-order transition but becomes second order at a tricritical point above 100 K.  These mean-field predictions for the low-field phase behavior of Er166 are in agreement with magnetization measurements \cite{Dhakal2021,Riberolles2024b}.

For planar fields, the evolution of the spiral and conical phases is quite complex. Studies of the evolution of the double-spiral phase in YMn$_6$Sn$_6$ reveals transverse conical spiral and fan-like phases \cite{Ghimire2020}.  These long-period distorted structures roughly occur in the hatched region of Fig.~\ref{fig:Er} and are outside of the scope of our constrained mean-field model and are a subject of future study. 

Here, we consider the evolution of the co-planar phases that appear at temperatures below $T_{spiral}$, but more generally at field strengths $\gtrsim 5-10$ T and are plotted in Appendix B Fig.~\ref{fig:R166_mag}(e). Starting with the field along the planar easy axis, ${\bf H}\parallel b$, a sequence of first-order transitions occurs. The FIM$b$ phase is stable up to 18 T at low temperatures before entering the HP($-$) phase where the Er moment flops into the next minimum in the six-fold planar anisotropy at $\Delta\varphi^R\approx\pm60\degree$. As another $\Delta\varphi^R$ jump into the HP($+$) phase costs too much exchange energy, the HP($-$) phase has only a narrow range of stability before switching to the VP$b$($\pm$) phase at 24 T where $\theta^R\sim 90\degree$. At even higher fields of 48 T, the canting plane switches again, this time into the HP($+$) phase after making the next $\Delta\varphi^R$ jump.  From here, the FF phase is achieved at 72 T by rotation of the Er and Mn moments towards the field. This switching of the canting plane is driven by the complex interaction between competing Er and Mn MAE and the exchange energy.  First-order FIM$b$-to-HP($-$) and HP($-$)-to-VPb($-$) transitions have both been observed in high field magnetization data at $\sim$20 T and $\sim$27 T, respectively \cite{Suga2006}.

At higher temperatures, the Er MAE evolves from weakly uniaxial to easy-plane which favors the HP-canted phase. Therefore, the VP-canted phases are squeezed out above $T=45$ K. At the same time, the Er planar anisotropy is strongly quenched and the pinning of the Er moment to six-fold planar MAE minima is suppressed.  This allows the system to evolve continuously from FIM$b$ to FF through the HP canted phase via two second-order transitions, similar to Gd166. Complete quenching of the Er anisotropy eventually squeezes out the HP-canted phase near $T_c^R=75$ K. The nested canted phases are centered along the $H^R_{\parallel}(T)$ line where the component of the Er moment parallel to the field vanishes.

For the field ${\bf H}\parallel a$ along the planar hard axis, the magnetization is not initially parallel to the field. At low-fields, there is a rotation of the net magnetization towards the field. Although the coherent rotation is not perfect, resulting in a HP($-$) phase. With increasing field, it is initially easier to switch into the VP$a$($\pm$) phase at 22 T (as observed in Ref.~\cite{Suga2006}) before the Er moment planar angle can jump into the next planar MAE minimum into the HP($+$) phase at 50 T. The critical field to enter the FF phase (87 T) is larger than ${\bf H}\parallel a$ due to the added planar anisotropy that must be overcome to fully align the moments with the field.  At temperatures above 45 K, the VP phases are squeezed out and we encounter a continuous transition from FIM$b$ to FIM$a$ via the HP($-$) phase (with nearly coherent ferrimagnetic rotation) with increasing field. Further field increases return to back to the HP phase.  Higher temperatures encounter the aforementioned triple-spiral phase with complex field evolution up to $\sim$10 T.

\subsection{TmMn$_6$Sn$_6$}

\begin{figure}[ht]
\includegraphics[width=1\linewidth]{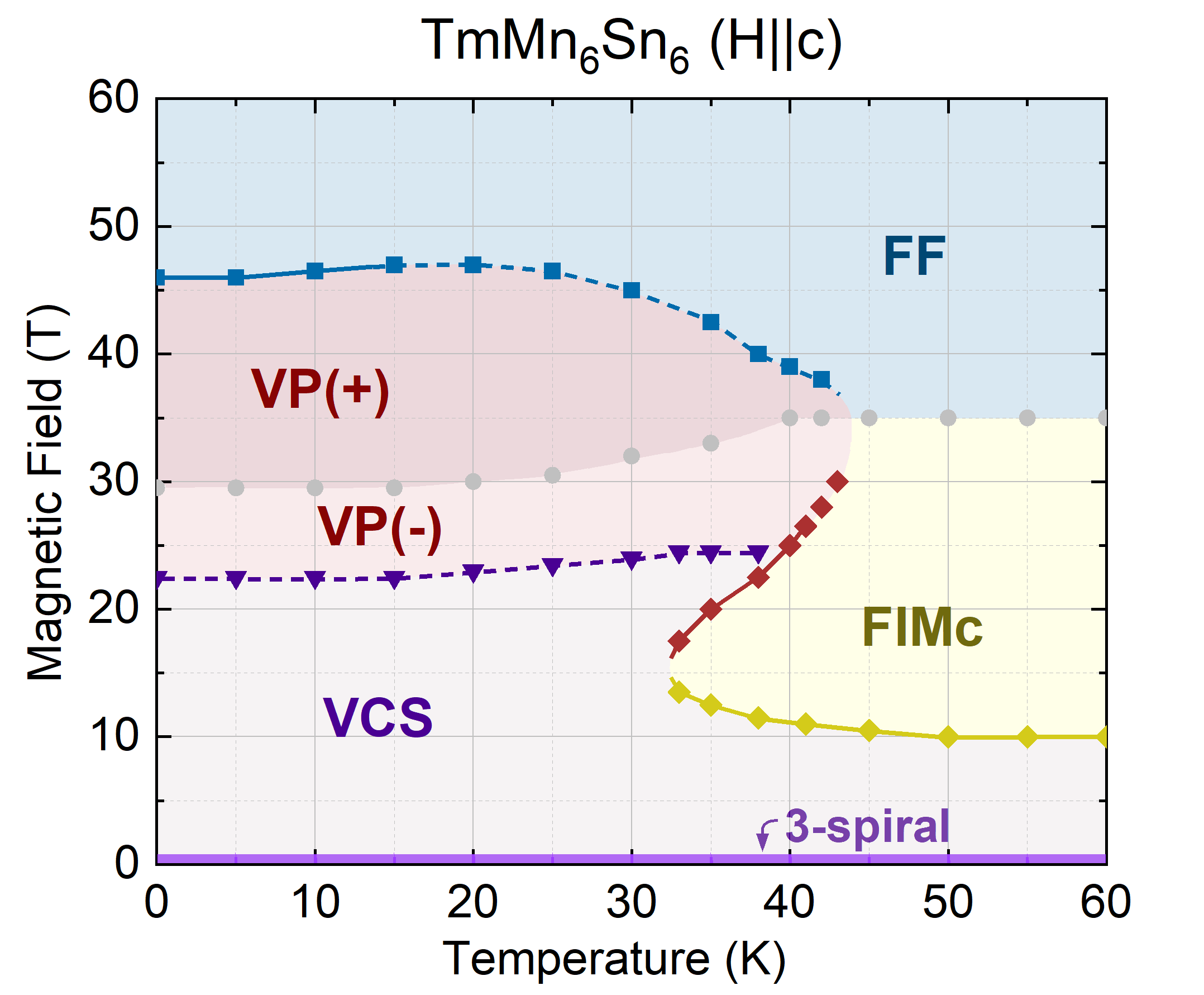}
\caption{Mean-field magnetic phase diagram for TmMn$_6$Sn$_6$ for an applied field along the $c$ direction. Labels for the different magnetic phases are defined in Table \ref{tbl:phases}. Colored symbols correspond to phase transitions determined from mean-field calculations. Solid (dashed) lines correspond to first-order-like (second-order-like) phase transitions, respectively. Gray circles mark the crossover lines corresponding to $H^R$ and $H_{\parallel}^R(T)$.}
\label{fig:Tm}
\end{figure}

Tm166 has the weakest Mn-$R$ exchange coupling and forms a zero-field chiral triple-spiral magnetic ground state as observed by neutron diffraction \cite{Lefevre2002}. Calculations of the ground state spiral period ($\Phi=48\degree$) and its temperature dependence (see Fig.~\ref{fig:R166_SR}(b) in Appendix B) are consistent with neutron-diffraction data up to $\sim$200 K.  Above 200 K, Tm166 evolves strongly towards collinear Mn-Mn antiferromagnetism ($\Phi = 180\degree$). This is not observed in our calculations and suggests that the effective Mn-Mn interlayer interactions become strongly temperature dependent close to $T_N = 325$ K.

The phase diagram for Tm166 with ${\bf H} \parallel c$ is shown in Fig.~\ref{fig:Tm}. In our model, application of a field with ${\bf H} \parallel c$ at low temperatures creates a VCS phase. However, the strong planar anisotropy of Tm results in a rather unusual configuration where Mn moments form a conical spiral, but the Tm moment remains in the basal plane and forms a simple spiral. As the field increases, the magnetization grows smoothly [see Appendix B Fig.~\ref{fig:R166_mag}(f)]. However, calculations show that the period of the conical spiral grows until it becomes indistinguishable from the VP$b$ canted phase above 22 T. This novel phase transition is unique among the $R$166 compounds and leads to a field-driven loss of spin chirality with a concomitant change of the MSG. It has no signature in the magnetization and would require high-field neutron diffraction to detect.

Continuing to increase the field leads to a first-order spin-flip transition into the FF phase at 46 T.  Here the Tm moment orients discontinuously from the basal plane to the $c$-axis field direction.  At higher temperatures, a first-order transition from the VCS phase into the FIM$c$ phase occurs with reentrant spin-flop like transitions into either the VP($-$) or VCS phases.  Similar to Er166, in-plane magnetic fields create distorted long-period noncollinear or noncoplanar structures that are outside the methods presented here.

\section{Conclusions}
R166 compounds were originally studied for their complex magnetic structures and corresponding field and temperature driven spin-reorientation transitions.  Here, we give a complete description of the many different magnetic phases that are possible in these compounds as a consequence of competing anisotropies and exchange. The temperature dependence is largely controlled by thermal fluctuations of the $R$ moment which act to quench the magnetic anisotropy and weaken the effective Mn-$R$ exchange coupling.  At high temperatures, this quenching favors planar collinear or spiral phases depending on the effective strength of the Mn-$R$ exchange. At intermediate temperatures and fields, this leads to a variety of spin-reorientation transitions between collinear phases and spin-flop and spin-flip transitions between collinear and canted phases.  These can be first or second-order transitions dependent on the detailed magnetic anisotropy landscape. Novel phase transitions corresponding to rotations of the canting plane are driven by competing exchange and anisotropic terms. Some transitions possess unique bicritical or tritcritical points or serve as novel magnetic analogs of liquid-gas transitions terminating at a critical point.

The nature of these phases and their magnetic symmetry serves as a starting point to study the effect of magnetism on band topology. It is clear that most spin-reorientation and spin-flop phases will change the magnetic space group which results in the breaking of different crystallographic symmetries. Some of these transitions have critical fields and temperatures that are accessible with conventional magnets, although access to many transitions require specialized high-field facilities. 

A major question is what relevance these magnetic symmetries may have on the electronic states and topology. Experimental work along these lines has considered the connection between longitudinal and Hall transport in R166 compounds at low fields \cite{Ghimire2020, Ma2021, Dhakal2021, Fruhling2024}. The observation of the topological Hall effect in Y166, Er166, and Tm166 \cite{Ghimire2020,Wang2022,Fruhling2024} suggests the development of scalar spin chirality. This includes the observation of skyrmion-like textures in thin films of Tb166 near the spin-reorientation transition \cite{Li2023}.  Here, the constraints of our mean-field model only allow us to find triple-spiral and VCS phases with nonzero vector chirality. As shown in Fig.~\ref{fig:R166_SR}(b), these phases appear in our calculations for Ho166, Er166, and Tm166 when the the effective $R$-Mn exchange is weak enough and the effective anisotropy is planar. The topological Hall effect is linked to the distortion of spiral and conical phases in planar magnetic fields into noncoplanar structures. Future experimental and theoretical work will consider the evolution of noncoplanar phases and spin chirality that appear in $R$Mn$_6$Sn$_6$ compounds.  

\section{Acknowledgments} This work is supported by the Center for the Advancement of Topological Semimetals (CATS), an Energy Frontier Research Center funded by the USDOE Office of Science, Office of Basic Energy Sciences, through the Ames Laboratory. Ames Laboratory is operated for the USDOE by Iowa State University under Contract No. DE-AC02-07CH11358. 

\section{Data Availability} The data that support the findings of this article are openly available \cite{data}.

\section{APPENDIX A: ANALYTICAL AND NUMERICAL RESULTS FOR ISOTROPIC IONS}
\subsection{Derivation of $H_c^{\pm}$}
The critical fields $H_c^{\pm}$ can be derived for an isotropic system at zero temperature by minimizing the energy of the $R$ and Mn sublattices in a magnetic field.  For ${\bf H}\parallel c$, the total energy per formula unit is
\begin{align}
    E = &-\mu_B\mu_0 H[6gs\cos\theta^M+g_RJ\cos\theta^R] \\ \nonumber
    &+12\mathcal{J}^{MR}Ss\cos(\theta^M+\theta^R). \\ \nonumber
\end{align}
In the equilibrium state the total magnetization perpendicular to the field is zero, leading to the constraint
\begin{equation}
    \sin\theta^R = \frac{6gs}{g_RJ}\sin\theta^M.
\end{equation}
We now consider the situation where ramping the field first leads to an instability between the FIM and canted phase at $H_c^-$.  We assume the angles are small and therefore $\theta^M=\delta$ and $\theta^R=\pi-\frac{6gs}{g_RJ}\delta$.  The total energy can be written to second-order in $\delta$ as
\begin{align}
    E=E_0& + \mu_B\mu_0H\Big[6gs-\frac{(6gs)^2}{g_RJ}\Big]\frac{\delta^2}{2} \\ \nonumber
    &+12\mathcal{J}^{MR}Ss\Big(1-\frac{6gs}{g_RJ}\Big)^2\frac{\delta^2}{2} \\ \nonumber
\end{align}
where $E_0 = -\mu_B\mu_0H[6gs-g_RJ]-12\mathcal{J}^{MR}Ss$ is the energy of the FIM phase.  
The transition to the canted phase occurs at the field $H_c^-$ where $dE/d\delta=0$.  This gives
 \begin{equation}
6g\mu_Bs\mu_0H_c^-\Big(1-\frac{6gs}{g_RJ}\Big)=-12\mathcal{J}^{MR}Ss\Big(1-\frac{6gs}{g_RJ}\Big)^2.
 \end{equation}
Using the relation that $S=(g_R-1)J$, this can be solved for $H_c^-$
\begin{equation}
    \mu_0H_c^-=\frac{2\mathcal{J}^{MR}(g_R-1)}{gg_R\mu_B}(6gs-g_RJ)
\end{equation}

A similar approach can be used to find $H_c^+$ near the boundary of the canted and FF phases.  Here the small angles are written as $\theta^M=\delta$ and $\theta^R=\frac{6gs}{g_RJ}\delta$.  The same approach gives
\begin{equation}
    \mu_0H_c^+=\frac{2\mathcal{J}^{MR}(g_R-1)}{gg_R\mu_B}(6gs+g_RJ)
\end{equation}

One can calculate the temperature dependence of $H_c^{\pm}$ by replacing $J$ with its temperature-dependent magnitude $\langle J\rangle$. For the isotropic case of Gd, the reduction of the moment follows a Brillouin function ($B_J$) obtained in the combined applied and molecular field acting on Gd.  The condition
\begin{equation}
    \langle J\rangle = JB_J\Big(\frac{g_R\mu_B J \mu_0(H_c^{\pm}-H^R)}{k_{B}T}\Big),
\end{equation}
where $H^R$ is the compensation field given in Eq.~\ref{eqn:comp_field}, must be self-consistently solved such that the value of $\langle J\rangle$ at the given temperature reproduces the correct critical fields.  The results for Gd166 are shown in Fig.~\ref{fig:Gd_analytical}.

\subsection{Derivation of $H_{\parallel}^R(T)$}
The $H_{\parallel}^R(T)$ line for Gd166 may be calculated using the condition that the net magnetization perpendicular to the applied field is zero while the Gd moment lies perpendicular to the applied field.  This leads to a constraint that $\sin\varphi_{\parallel}^M = g_R\langle J\rangle/6gs$ where $\varphi_{\parallel}^M$ is the canting angle of the Mn sublattice at this condition and $\langle J\rangle$ is the temperature-dependent magnitude of the Gd angular momentum.  The Mn canting causes a molecular-field component of magnitude $H_{\perp}^{\prime}=H^R\sin\varphi_{\parallel}^M$ that is perpendicular to the applied field. 

\begin{figure}[ht]
\includegraphics[width=0.9\linewidth]{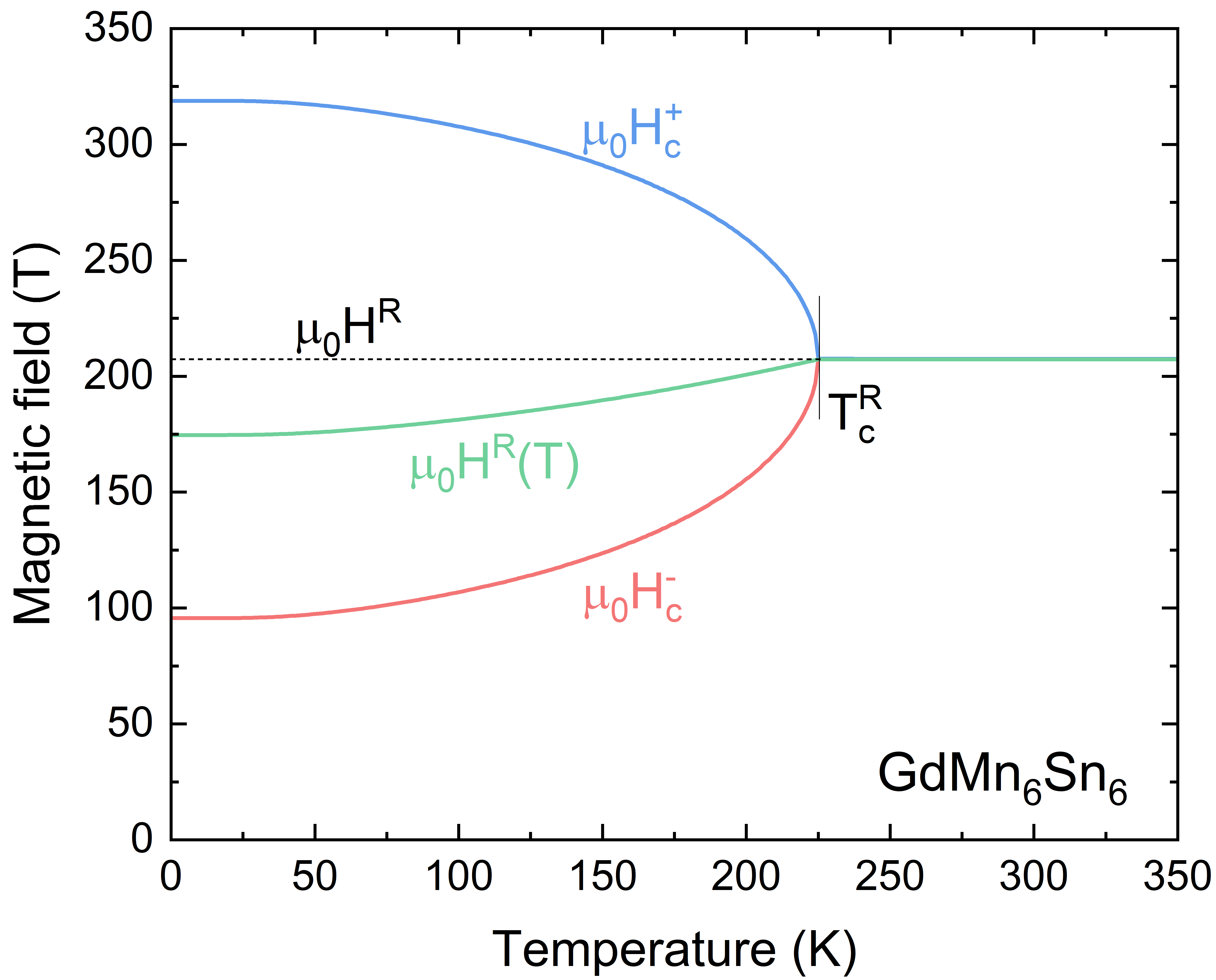}
\caption{Temperature dependence of the critical phase lines and crossover lines for GdMn$_6$Sn$_6$ calculated self-consistently using a numerical approach.
}
\label{fig:Gd_analytical}
\end{figure}

At zero temperature, $\langle J\rangle = J$ and it is easy to show that $H_{\parallel}^R(0) = H^R\cos\varphi^M_{max}$ as given in Eq.~\ref{eqn:HR0}. At finite temperature, the magnitude of the Gd moment is reduced as given by the Brillouin function 
\begin{equation}
    \langle J\rangle = JB_J\Big(\frac{g_R\mu_B J \mu_0H_{\perp}^{\prime}}{k_{B}T}\Big).
\end{equation}
This relation must be self-consistently solved for $\sin\varphi_{\parallel}^M$.  Numerical calculations were performed to obtain $\sin\varphi_{\parallel}^M$ [and $H_{\parallel}^R(T)$] for any temperature.  Self-consistent numerical calculations of $H_c^{\pm}$ and $H_{\parallel}^R(T)$ shown in Fig.~\ref{fig:Gd_analytical} are in perfect agreement with the results from mean-field energy minimization shown in Fig.~\ref{fig:Gd}.

\subsection{Parametrization of Mn canting in GdMn$_6$Sn$_6$}

For the isotropic GdMn$_6$Sn$_6$ system, we can parametrize the Mn canting angle at any temperature and field within the canted phase region by using the expression
\begin{equation}
    \cos\varphi^M = \frac{B(H,T)(H-H^R)^2-A(H,T)}{H}-1
\end{equation}
where $A$ and $B$ are temperature and field dependent parameters.  Using the condition that $\varphi^M=0$ at $H=H_c^-$, it is straightforward to show that
\begin{equation}
    \frac{A}{B} = \mu_0^2(H_c^- - H^R)^2= (\mu_0H^R)^2\Big(\frac{g_R\langle J \rangle}{6gs}\Big)^2
\end{equation}
Next, we use the condition that $\varphi^M=\varphi^M_{\parallel}$ along the line $H=H_{\parallel}^R(T)$.  $\varphi^M_{\parallel}$ is determined numerically as described in the previous section. This gives
\begin{equation}
    A = \mu_0H^R\frac{\cos\varphi^M_{\parallel}\sin^2\varphi^M_{\parallel}(1-\cos\varphi^M_{\parallel})}
{\sin^2\varphi^M_{\parallel}-(1-\cos\varphi^M_{\parallel})^2}
\end{equation}

\section{APPENDIX B: SELECT MEAN-FIELD MAGNETIZATION CALCULATIONS} \label{sec:App_mag}
\begin{figure}[ht]
\includegraphics[width=0.7\linewidth]{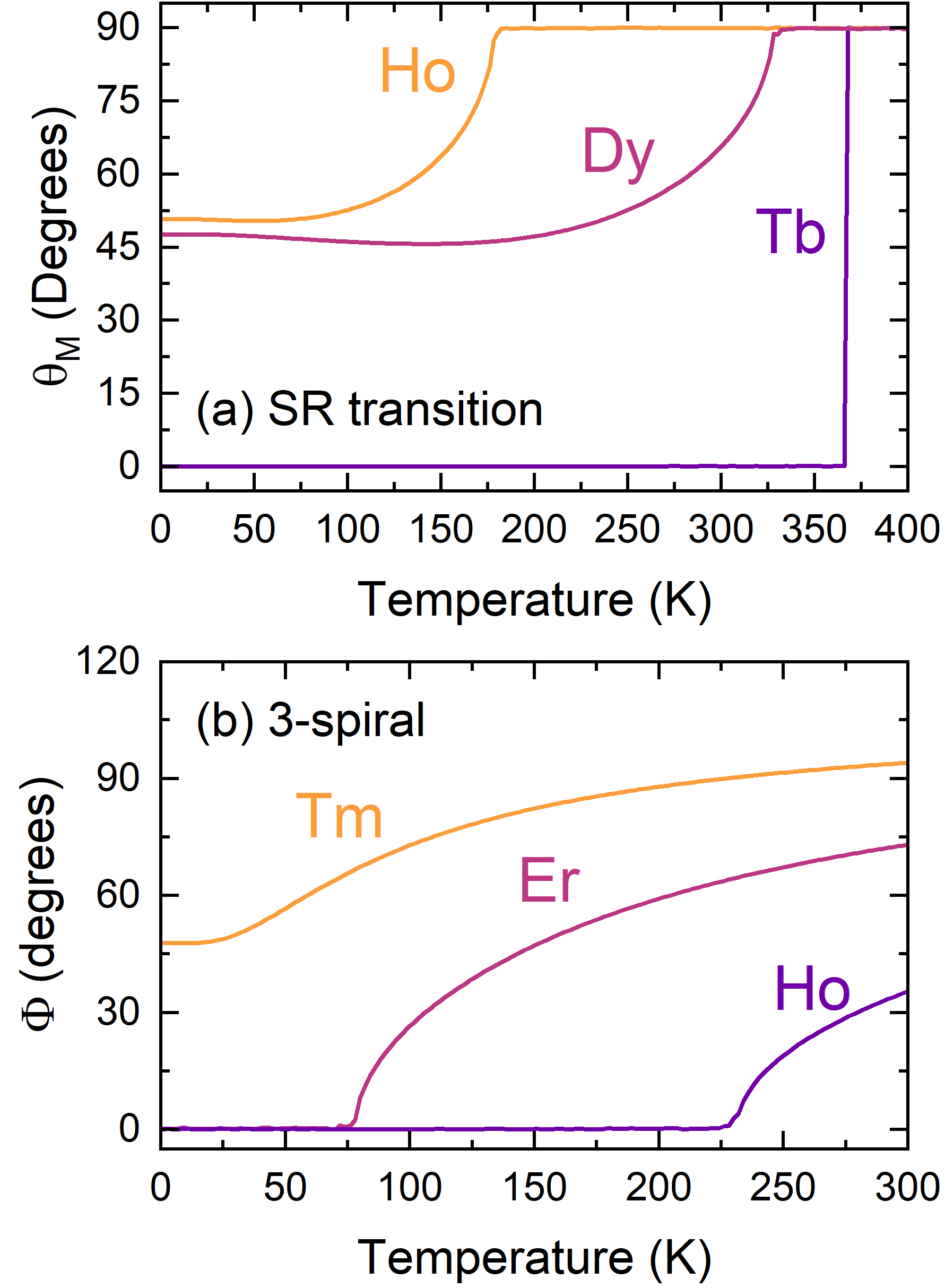}
\caption{Mean-field calculations of the zero-field temperature dependence of (a) the polar angle of Mn spin for Tb166, Dy166, and Ho166 showing the spin-reorientation transition, and (b) the helical pitch angle for Ho166, Er166, and Tm166.
}
\label{fig:R166_SR}
\end{figure}

\begin{figure}[ht]
\includegraphics[width=1\linewidth]{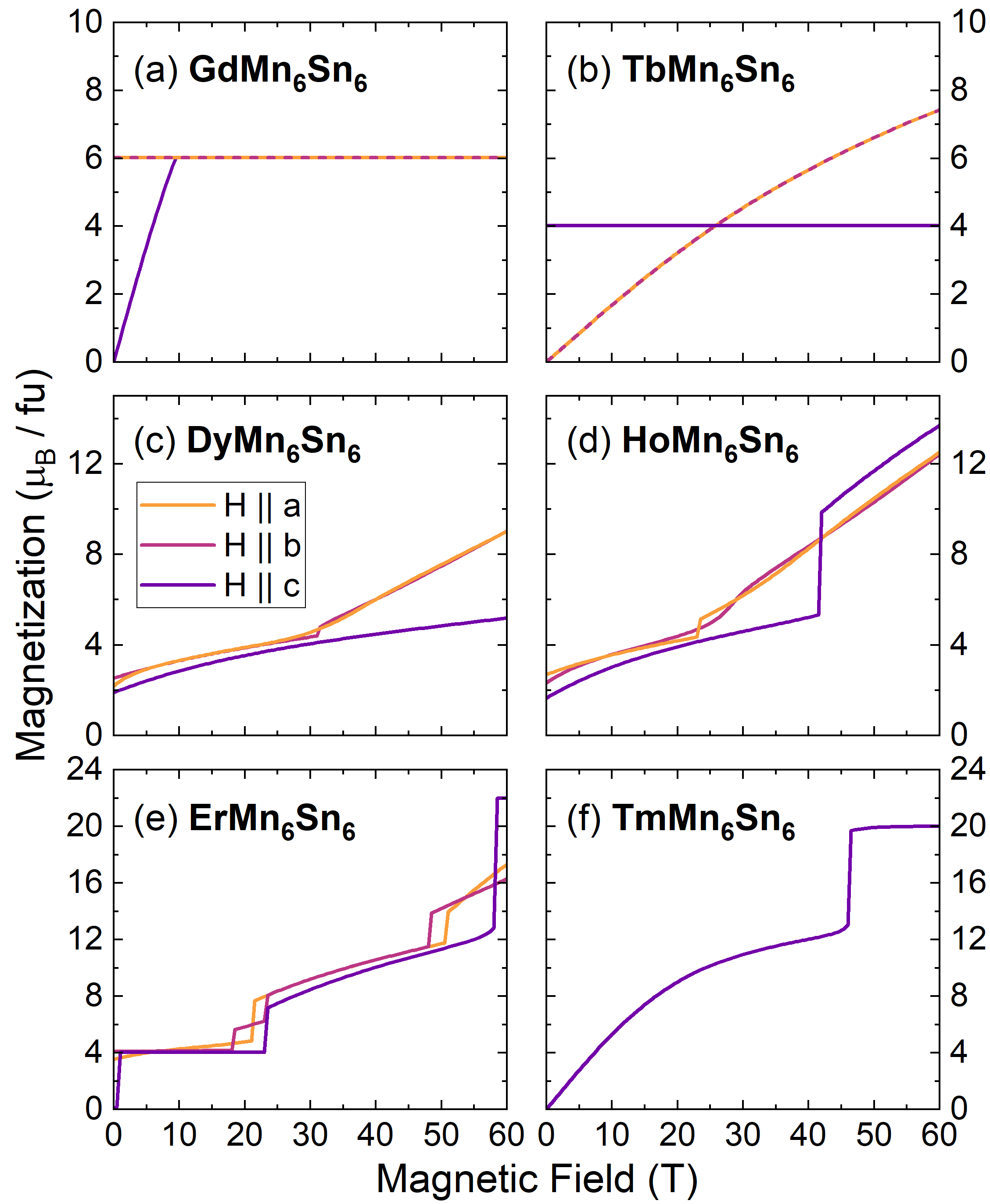}
\caption{(a)-(f) Mean-field calculations of the low-temperature ($T = 0.1$~K) magnetization for $R$166 compounds with the field applied along $a$, $b$, or $c$ directions, as defined in Fig.~\ref{fig:structure}.
}
\label{fig:R166_mag}
\end{figure}

Figure~\ref{fig:R166_SR} plots mean-field calculations of the zero-field spin-reorientation transitions and helical transitions as a function of temperature.  These calculations can be compared to experimental data in Refs.~\cite{Malaman1999, Lefevre2002, Riberolles2024b}. There are no observations of a helical phase in Ho166 at high temperatures.  Figure \ref{fig:R166_mag} plots mean-field calculations of the low-temperature magnetization of each $R$166 compound for different field directions.  These results can be compared with high-field magnetization data in Refs.~\cite{Kimura2006, Suga2006, Sawai2006}.

\bibliography{R166}
\end{document}